\documentclass[useAMS,usenatbib]{mn2e}

\usepackage{graphicx}
\usepackage{amssymb}

\newcommand{\new}[1]{{\rm #1}}
\defcitealias{Lane09}{Paper I}
\defcitealias{Lane10a}{Paper II}

\title[Halo Globular Clusters]{Halo Globular Clusters Observed with
  AAOmega:\\Dark Matter Content, Metallicity and Tidal Heating}
  \author[Richard R.  Lane {\it et al.}]{Richard
  R.  Lane$^{1}$\thanks{E-mail: rlane@physics.usyd.edu.au}, L\'aszl\'o
  L.  Kiss${^{1,2}}$, Geraint F.  Lewis${^1}$, Rodrigo
  A. Ibata${^3}$,\vspace{2.5mm}\\
\hspace{-1mm}{\LARGE    {\rm   Arnaud   Siebert${^3}$,    Timothy   R.
    Bedding${^1}$,       P\'eter       Sz\'ekely$^{4}$,       Zolt\'an
    Balog$^{5}$}}\vspace{2.5mm}\\
\hspace{-1mm}{\LARGE {\rm and Gyula M. Szab\'o${^2}$}}\\ $^{1}$Sydney
    Institute for Astronomy, School of Physics, A28, The University of
    Sydney, NSW, Australia 2006\\ $^{2}$Konkoly Observatory of the
    Hungarian Academy of Sciences, PO Box 67, H-1525, Budapest,
    Hungary\\ $^{3}$Observatoire Astronomique, Universite de
    Strasbourg, CNRS, 67000 Strasbourg, France\\ $^{4}$Department of
    Experimental Physics, University of Szeged, Szeged 6720, Hungary\\
    $^{5}$Max-Planck-Institut f\"ur Astronomie, K\"onigstuhl 17,
    D-69117 Heidelberg, Germany\\}
\begin{document}

\date{Accepted for publication in MNRAS}

\pagerange{\pageref{firstpage}--\pageref{lastpage}} \pubyear{2009}

\maketitle

\label{firstpage}

\begin{abstract}
Globular clusters have proven to be essential to our understanding of
many important astrophysical phenomena. Here we analyse spectroscopic
observations of ten Halo globular clusters to determine their dark
matter content, their tidal heating by the Galactic disc and halo,
describe their metallicities and the likelihood that Newtonian
dynamics explain their kinematics.  We analyse a large number of
members in all clusters, allowing us to address all these issues
together, and we have included NGC 288 and M30 to overlap with
previous studies.  We find that any flattening of the velocity
dispersion profiles in the outer regions of our clusters can be
explained by tidal heating.  We also find that all our GCs have
M/L$\rm _V\lesssim5$, therefore, we infer the observed dynamics do not
require dark matter, or a modification of gravity.  We suggest that
the lack of tidal heating signatures in distant clusters indicates the
Halo is not triaxial.  The isothermal rotations of each cluster are
measured, with M4 and NGC 288 exhibiting rotation at a level of
$0.9\pm0.1$\,km\,s$^{-1}$ and $0.25\pm0.15$\,km\,s$^{-1}$,
respectively. We also indirectly measure the tidal radius of NGC 6752,
determining a more realistic figure for this cluster than current
literature values. Lastly, an unresolved and intriguing puzzle is
uncovered with regard to the cooling of the outer regions of all ten
clusters.
\end{abstract}

\begin{keywords}
gravitation - Galaxy: globular clusters: individual - stellar dynamics
\end{keywords}

\section{Introduction}

Globular clusters (GCs) are often used as tracers of the gravitational
potentials of galaxies and galaxy clusters
\cite[e.g.][]{Kissler-Patig99,Cote03,Wu06,Quercellini08,Gebhardt09}. Although
this has been applied to theoretical Galactic potentials
\cite[e.g.][]{Allen06}, the actual Milky Way (MW) potential has not
yet been analysed in this way. The tidal forces of spiral galaxies are
thought to be strongest near the disc because the concentrated mass in
that region (gas and stars) has a larger density gradient than the
more slowly varying density of the dark matter (DM) halo.
Interestingly, many distant MW objects such as GCs and dwarf galaxies
are known to be tidally stripped, despite being far enough from the
Disc that they should not directly interact with it.  For example, NGC
7492 is $\sim3$\,kpc further from the Galactic centre than the most
distant detection of the Monoceros Ring, an object on the very
outskirts of the Disc \cite[$\sim22$\,kpc;][]{Conn07}, and exhibits
clear evidence of tidal interaction with the Galaxy \cite[][]{Lee04}.
In this paper we consider 10 GCs at varying Galactocentric and Planar
distances, allowing the inference of properties of the potential of
the MW for the first time, by looking for signatures of tidal heating
in these clusters.

Many globular clusters exhibit internal accelerations below
$a_0\approx1.2\times10^{-10}$\,m\,s$^{-2}$, the level at which either
modified gravity \cite[e.g. MOND;][]{Milgrom83} or dark matter
is required to reconcile the observed kinematics of elliptical
galaxies with theory. Near the tidal radius ($r_t$) it is likely that
most stars in GCs feel accelerations below this level, making them an
ideal testing ground for low-acceleration gravity \cite[][and
references therein]{Sollima10}.  Furthermore, if all GCs exhibit
similar behaviour, Galactic influences can not be the primary cause.
In this final paper in the series [see also \citealt{Lane09}
(hereafter Paper I) and \citealt{Lane10a} (hereafter Paper II)], we
present the velocity dispersions and mass-to-light profiles of four
GCs, namely M4, M12, NGC 288 (chosen for comparison with earlier
studies) and NGC 6752, bringing the total for this project to 10. This
sample allows statistically significant conclusions to be made on the
dark matter content of Halo GCs, and on whether a modification of
gravity is required to reconcile their internal kinematics with
Newtonian gravitational theory.

Our sample of GCs contains three close to the Galaxy (M55, M12 and
M22), four at intermediate distances (NGC~6752, M4, M30 and 47~Tuc)
and three that are distant (M68, NGC~288 and M53).  We define `close'
to be $R<5$\,kpc, `intermediate' as $5<R<10$\,kpc, and `distant' to be
$R>10$\,kpc, following \cite{Harris96}.  NGC 288 was chosen, in part,
because it is one of the GCs analysed by \cite{Scarpa07b} who found it
to have a flat velocity dispersion profile, similar to that of
Low Surface Brightness galaxies which are thought to be DM dominated
through to their cores.  Our targets were then analysed in separate
studies (Papers I and II and the current paper) ensuring a mix of
nearby, intermediate and distant GCs to ensure any Galactic
influences, if any, would be clearly observed. See Table
\ref{metaltable} for the estimated acceleration, due to the cluster,
of the most distant cluster member for all ten clusters analysed in
this project. Note that the three distant clusters all experience
accelerations due to the Galaxy of $\sim a_0$.

\section[]{Data Acquisition and Reduction}\label{data}

We used AAOmega, a double-beam, multi-object spectrograph on the 3.9m
Anglo-Australian Telescope (AAT) at Siding Spring Observatory in New
South Wales, Australia, to obtain the data for this survey.  AAOmega
covers a two-degree field of view, and is capable of obtaining spectra
for 392 individual objects over this field. We used 30 sky fibres used
for optimal sky subtraction, and 5--8 fibres for guiding.  The
positional information for our targets was taken from the 2MASS Point
Source Catalogue \citep{Skrutskie06} which has an accuracy of
$\sim0.1''$.

Observations of M4 were performed on February 15--17, 2008, with
$1.5''-2.5''$ seeing.  The data for M12 were taken over two observing
runs: 7 nights on August 12--18 2006, and a further 8 nights on August
30 -- September 6 2007, both with mean seeing of $\sim1.5''$. NGC 288
was observed during the 2006 run and NGC 6752 during the 2007 run.
For all observations we used the 2500V grating in the blue arm,
resulting in spectra between $4800{\rm \AA}$ and $5150{\rm \AA}$ with
$\lambda/\Delta\lambda = 8000$.  In the red arm we used the 1700D
grating, which is optimized for the CaII IR triplet region.  The red
spectra cover $8350-8790{\rm \AA}$, with $\lambda/\Delta\lambda =
10000$.  This setup returns the highest spectral resolution available
with AAOmega, and is suitable for measuring stellar radial velocities.
We selected targets for this campaign by matching the $J-K$ colour and
$K$ magnitude range of the red giant branch (RGB) of each cluster.  To
minimize scattered-light cross-talk between fibres, each configuration
was limited to 3 magnitudes in range.

We obtained 718, 2826, 1223 and 3664 spectra in the M4, M12, NGC 288
and NGC 6752 fields, respectively.  Flat-field and arc-lamp exposures
were used to ensure accurate data reduction and wavelength
calibration.  Data reduction was performed with the {\tt2dfdr}
pipeline\footnote{http://www2.aao.gov.au/twiki/bin/view/Main/CookBook2dfdr},
which was specifically developed for AAOmega data. We checked the
efficacy of the pipeline with a comparison of individual stellar
spectra.

Radial velocities and atmospheric parameters were obtained through an
iterative process, taking the best fits to synthetic spectra from the
library by \cite{Munari05}, degraded to the resolution of AAOmega, and
cross-correlating this model with the observed spectra to calculate
the radial velocity [a process very similar to that used by the Radial
Velocity Experiment \cite[RAVE;][]{Steinmetz06,Zwitter08} project].
We used the same spectral library as the RAVE studies; this process is
outlined in detail by \cite{Kiss07}.

\subsection{Cluster Membership}\label{Cluster Membership}

We determined cluster membership using four parameters: the equivalent
width of the calcium triplet lines, surface gravity, radial velocity
and metallicity ([m/H]).  Stars matching all criteria were judged to
be members. Only stars having $\log g<4.0$ and $\log g<4.6$ were
selected for NGC 6752 and M12, respectively, ensuring the majority of
Galactic contaminants were removed before further selection criteria
were applied.  This probably removed some genuine cluster members but
was necessary to ensure our sample was as free from Galactic field
stars as possible.

\begin{figure*}
  \begin{centering}
  \includegraphics[angle=-90,width=0.48\textwidth]{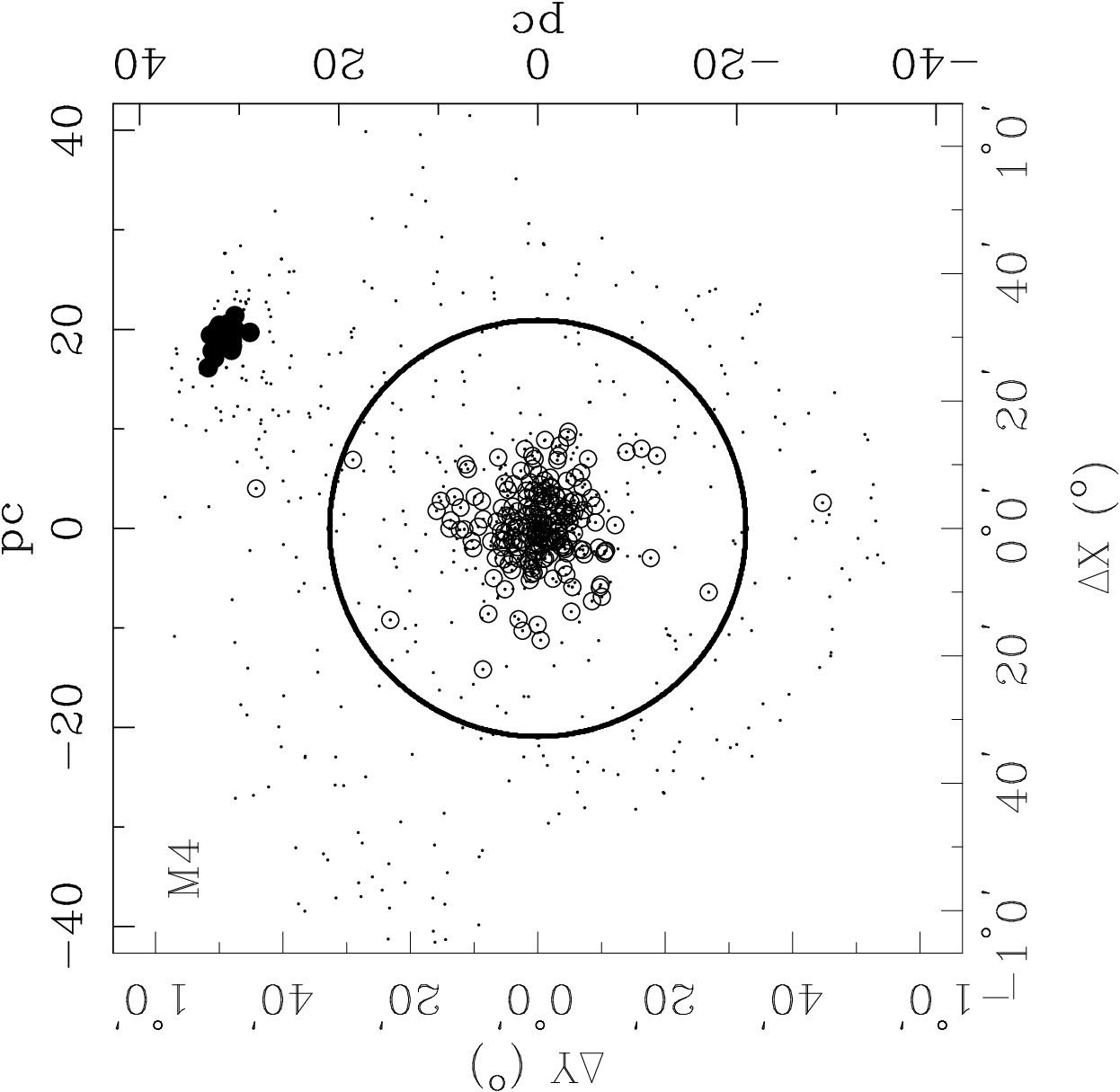}
  \includegraphics[angle=-90,width=0.48\textwidth]{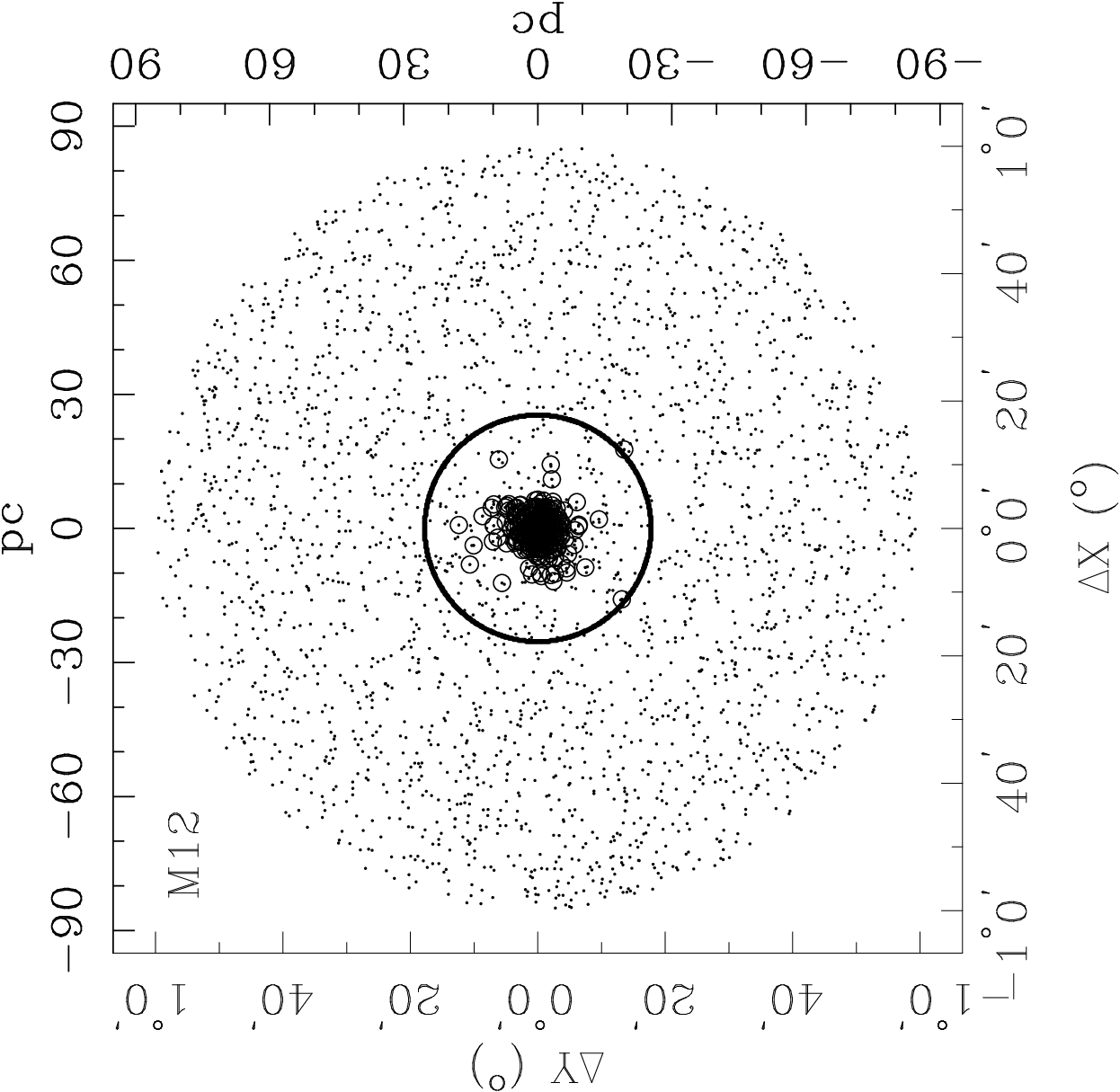}
  \includegraphics[angle=-90,width=0.48\textwidth]{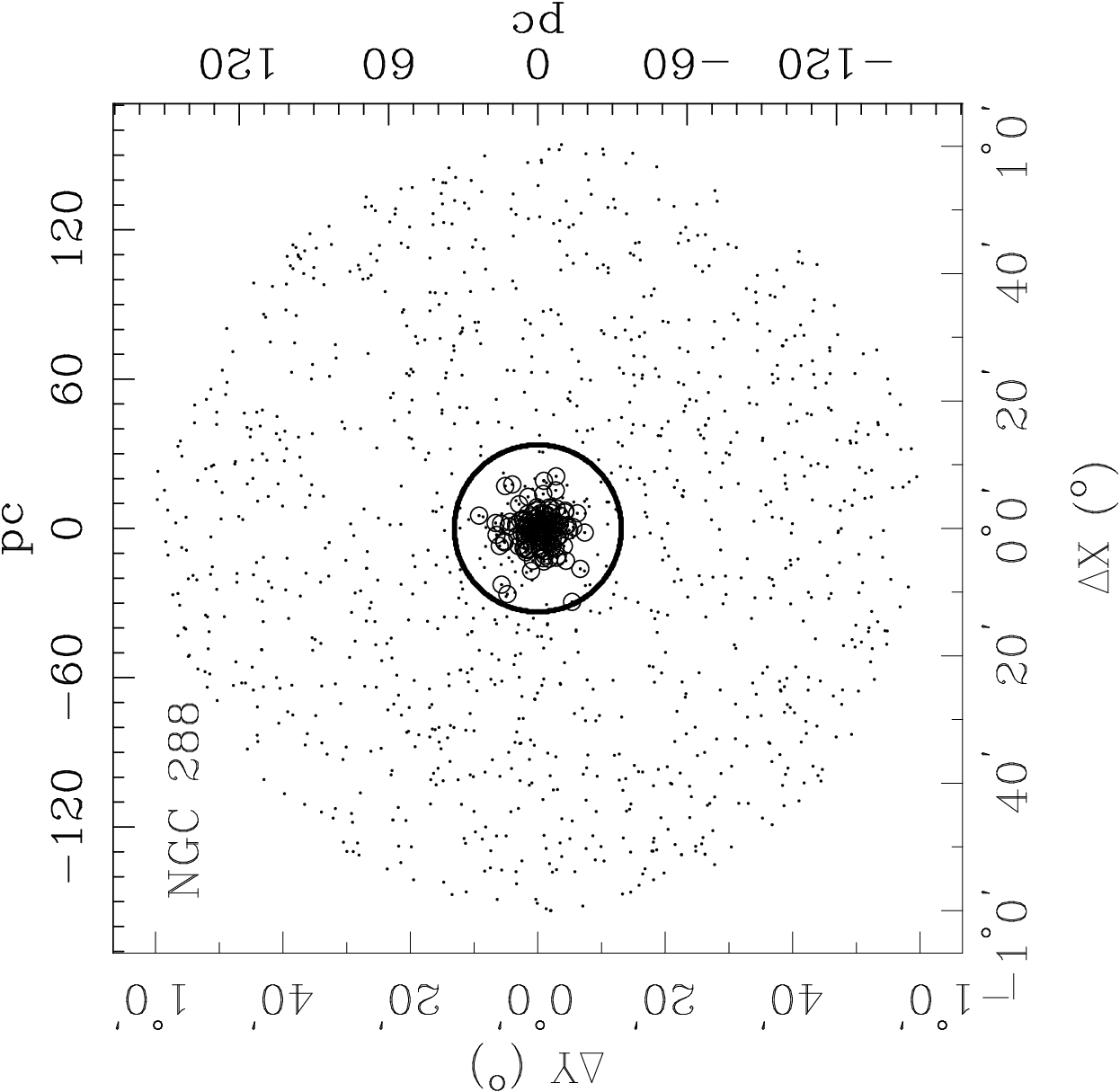}
  \includegraphics[angle=-90,width=0.48\textwidth]{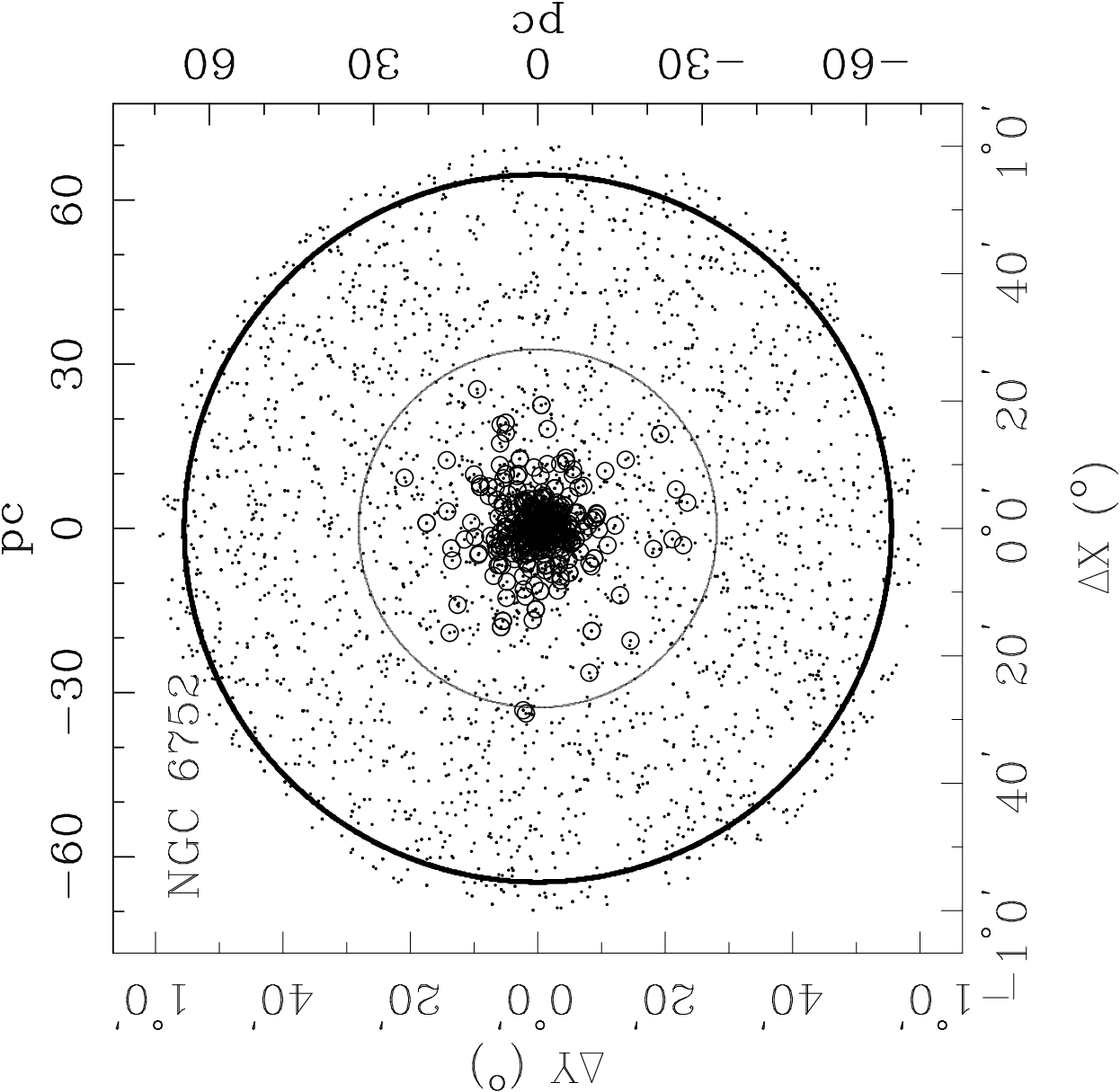}
  \caption{Distribution on the sky of the stars observed in the four
    fields, with axes in both degrees and parsecs from the
    cluster centre. Circled points indicate stars that we determined
    to be cluster members (see text).  The large circle is the tidal
    radius of the cluster from \citet{Harris96}, with the smaller
    thinner circle in the lower right panel being our derived value for
    $r_t$ for NGC 6752 (see text).  The large points in the upper
    right of the M4 field are the stars we determined to be members of
    NGC 6144.  In each panel, North is up and East is to the left.}
  \label{members}
  \end{centering}
\end{figure*}

For several clusters studied in \citetalias{Lane09}, a cutoff of
$T_{\rm eff}\gtrsim9000$\,K was necessary to remove hot horizontal
branch (HB) stars.  These have radial velocities with large
uncertainties due to the calcium triplet in very hot stars being
replaced by hydrogen Paschen lines.  No cuts were made on $T_{\rm
eff}$ for any of the current clusters because no stars with $T_{\rm
eff}\gtrsim7425$\,K (for M4), $T_{\rm eff}\gtrsim5600$\,K (for M12),
$T_{\rm eff}\gtrsim7000$\,K (for NGC 288) or $T_{\rm
eff}\gtrsim5500$\,K (for NGC 6752) remained after our selection
process.  In total, 200, 242, 133 and 437 stars were found to be
members of M4, M12, NGC 288 and NGC 6752, respectively.  Figure
\ref{members} shows the relative locations of the observed stars and
highlights those found to be members. Note that 19 stars in the M4
field were found to be members of the globular cluster NGC 6144. For
M4, we find that 88.0\% of the selected members fall within within
$2\sigma$ of $all$ selection parameters and 100\% within $3\sigma$.
For M12 these values are: 94.2\% and 100\%, for NGC 288: 89.5\% and
100\% and for NGC 6752: 97.5\% and 100\%. Based on this we see no
statistical reason to think there is significant Galactic
contamination in our final samples.

\section{Results}

\subsection{Tidal Radius of NGC 6752}\label{r_t6752}

The  tidal radius  of NGC  6752  in the  \cite{Harris96} catalogue  is
$55.\!'34$  (shown as the  large thick  circle in  the right  panel of
Figure \ref{members}), however,  this GC is known to  have a collapsed
core  \cite[e.g.][]{Rubenstein97}  and  \cite{Harris96} warns  against
using tidal  radii calculated from  core parameters for such  GCs. The
most  distant of our  selected members  throughout this  project have,
generally, been  very close to the  tidal radius. NGC 6752  is a clear
outlier, with members  only found to within $\sim1/2$  of the value of
$r_t$  quoted by  \cite{Harris96}.  Therefore,  we propose  an updated
value of $r_t$  for NGC 6752 based on  our membership selections. Note
that M30  and 47  Tuc were not  used to  determine $r_t$ for  NGC 6752
because of a paucity of stars  observed in M30, and because stars were
found out to the edge of the field of view of AAOmega for 47 Tuc.

Based on the membership selections of the remaining 7 clusters, the
tidal radius is located $94.1\pm2.1$\% of the distance to the most
distant member from the cluster centre.  Our most distant member for
NGC 6752 is located at $33.87$\,pc, or $29.\!'11$, therefore,
$r_{t_{6752}}=27.\!'4\pm1.\!'7$. The quoted uncertainty is based
only on the standard deviation of the distance between the outermost
member and the tidal radii of our clusters, so the true uncertainty in
the value of $r_t$ is likely to be much greater than this. The large
thin circle in the right hand panel of Figure \ref{members} represents
our derived value for $r_t$. It should be stressed that this value is
not a robust measure of $r_t$ but we suggest it is a more realistic
value than that in the \cite{Harris96} catalogue.

\subsection{Metallicity}\label{metaldist}

Our metallicity  ([Fe/H]) calibration  method, discussed in  detail in
\citetalias{Lane10a}, was used to  determine the metallicities of each
cluster.   Briefly, the  $K$ magnitude  of the  Tip of  the  Red Giant
Branch ($K_{\rm TRGB}$) was subtracted from the $K$ magnitudes of each
star and plotted  against the equivalent width of  the calcium triplet
lines to give  a distance independent measure of  luminosity.  For M4,
the  $K_{\rm TRGB}$ value  was taken  from a  $J-K$ versus  $K$ Colour
Magnitude  Diagram based  on 2MASS  data  within $5'$  of the  cluster
centre  ($K_{\rm  TRGB}=5.3$), for  M12  from \cite{Paust06}  ($K_{\rm
TRGB}=9.1$),  for NGC 288  from \cite{Davidge97}  and \cite{Valenti04}
($K_{\rm TRGB}=8.5$), and for  NGC 6752 from \cite{Valenti04} ($K_{\rm
TRGB}=7.4$).  Linear fits to these data, combined with plotting [Fe/H]
values  vs  $\Sigma  W  -  AX$   for  47  Tuc  and  M55  ([Fe/H]  from
\citealt{Harris96}, with $A$ being the gradient of the slope above and
$X$ being $K - K_{\rm TRGB}$), allows a calibrator on [Fe/H] for other
clusters.

\begin{figure}
  \begin{centering}
  \includegraphics[angle=-90,width=0.48\textwidth]{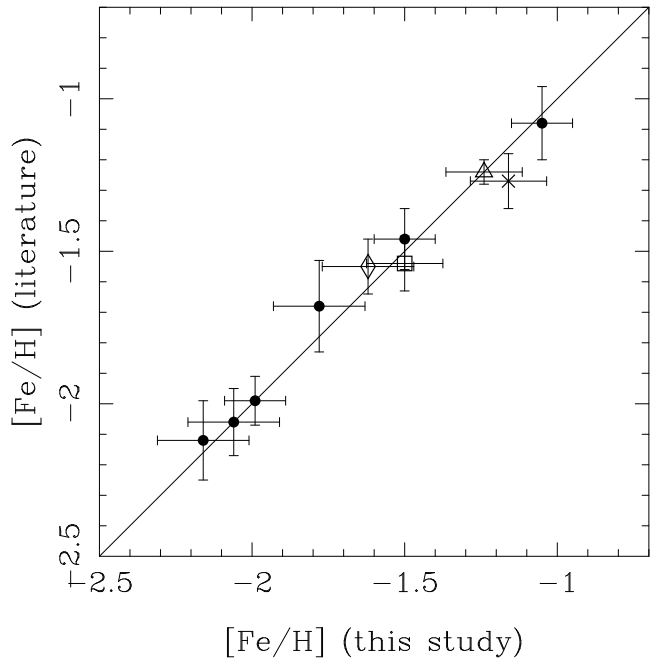}
  \caption{[Fe/H] values derived from our method outlined in the text
  versus those from the literature.  Solid points are those GCs from
  Papers I and II, and those from the current paper are: M4 (cross),
  M12 (square), NGC 288 (triangle) and NGC 6752 (diamond).  The [Fe/H]
  literature values are from \citet{Kanatas95}, \citet{Johnson06},
  \citet{Chen00} and \citet{Zinn85} respectively.}
  \label{metals}
  \end{centering}
\end{figure}

Figure \ref{metals} displays the robustness of this technique, with
[Fe/H] values from this project plotted against those from the
literature.  Solid points are the clusters analysed in Papers I and
II.  From the current paper, the cross is M4, the square is M12, the
triangle is NGC 288 and the diamond is NGC 6752; the [Fe/H] literature
values are from \cite{Kanatas95}, \cite{Johnson06}, \cite{Chen00} and
\cite{Zinn85} respectively. This method is similar to that outlined by
\cite{Cole04} and \cite{Warren09}, except we use the TRGB rather than
the HB so it can be used for much more distant objects. A recent
photometric study of 47 Tuc has revised the metallicity of 47 Tuc to
$-0.83$ \cite[][]{Bergbusch09}.  If this new value is adopted for our
calibration, a maximum change in our calculated [Fe/H] values is -0.05
(Kron 3), which is well within our uncertainty estimates. Calculated
[Fe/H] values are also shown in Table~\ref{metaltable}.

\subsection{Rotation}\label{rotation}

\begin{figure*}
  \begin{centering}
  \includegraphics[angle=-90,width=0.48\textwidth]{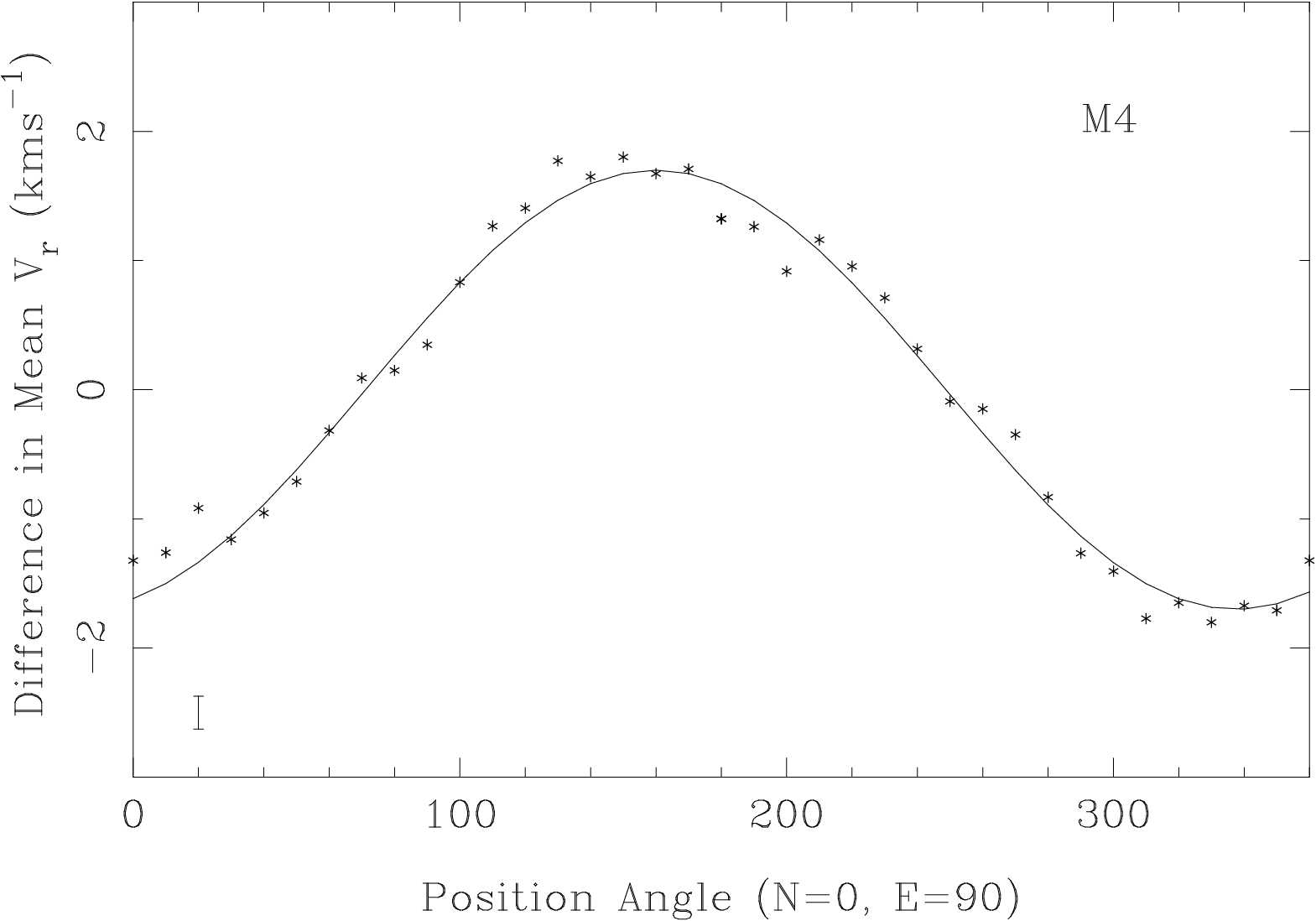}
  \includegraphics[angle=-90,width=0.48\textwidth]{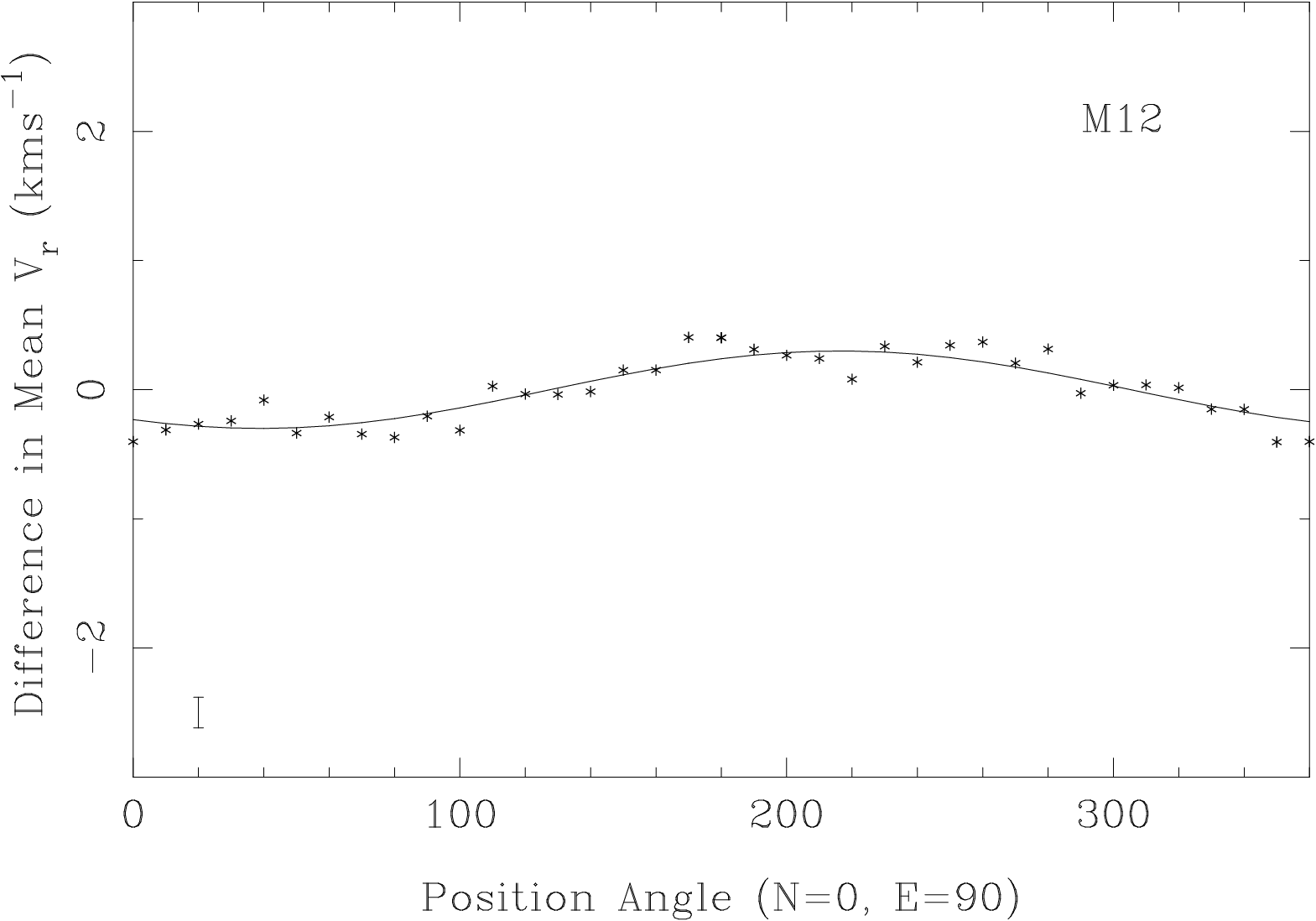}
  \includegraphics[angle=-90,width=0.48\textwidth]{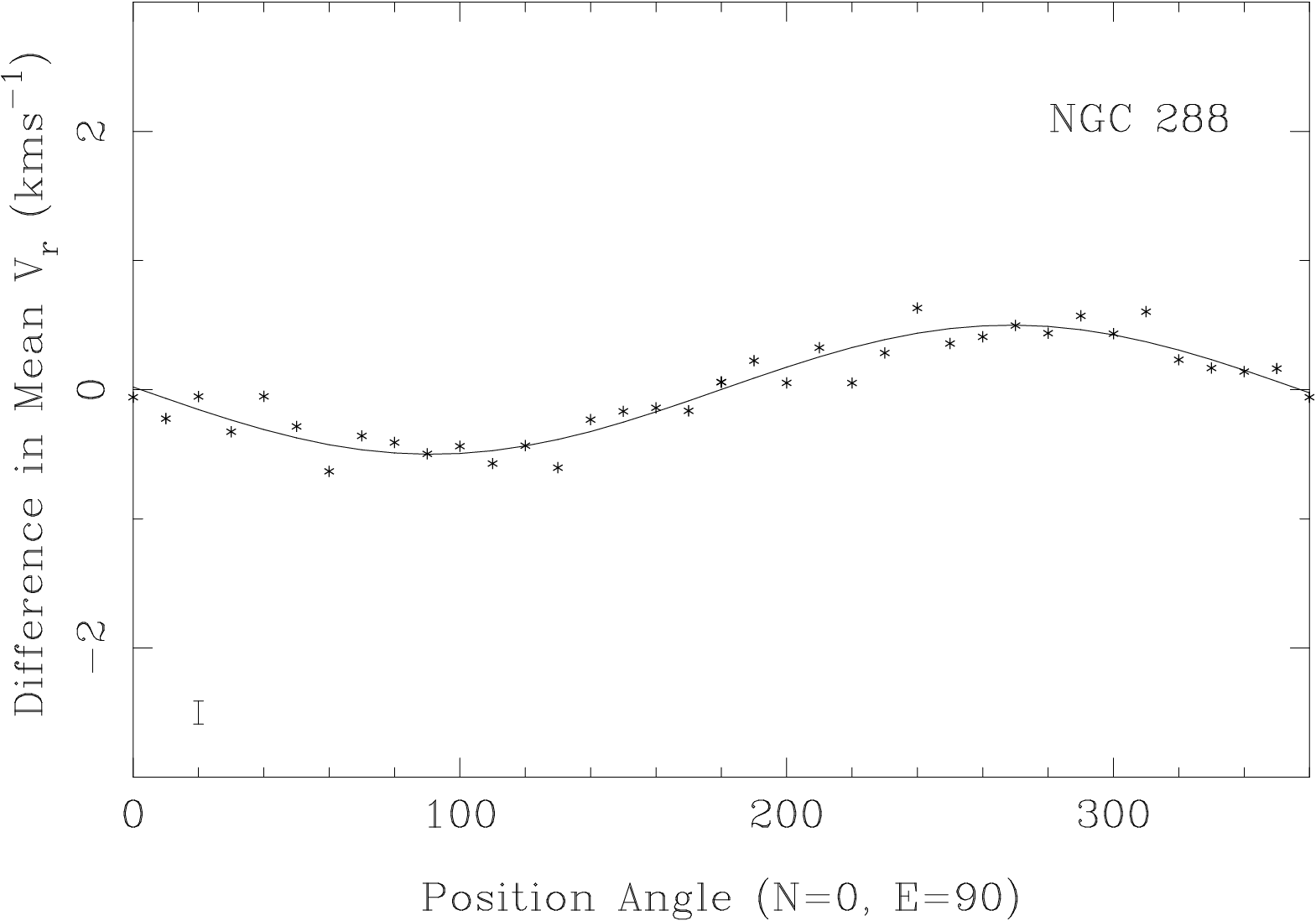}
  \includegraphics[angle=-90,width=0.48\textwidth]{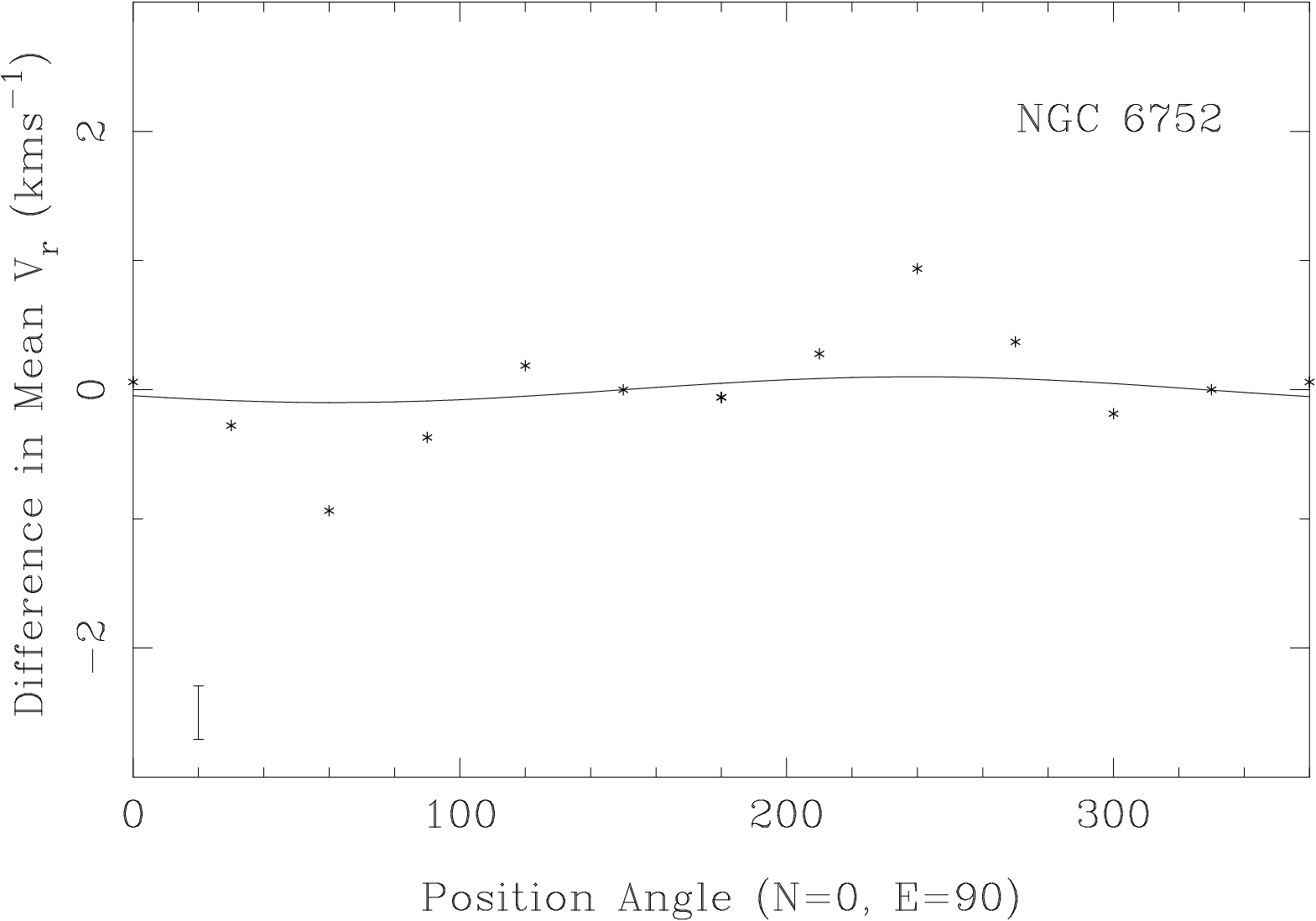}
  \caption{The rotation  of each cluster calculated  as the difference
    between  the mean  velocities on  each side  of the  cluster along
    equal position angles, as described in the text.  The best fitting
    sine  function  is  overplotted,   and  a  typical  error  bar  is
    represented in the lower left of each panel.}
  \label{rotationfig}
  \end{centering}
\end{figure*}

To  measure the  projected rotation  of  each cluster,  we assumed  an
isothermal distribution.  The rotations  were measured by halving each
by position  angle (PA) and  subtracting the mean stellar  velocity of
one half from the other. This  was repeated in steps of $10^\circ$ and
the     best-fitting     sine     function     overplotted     (Figure
\ref{rotationfig}). Note that for NGC 6752 it was necessary to perform
this in  steps of  $30^\circ$ to avoid  aliasing effects.   The method
results  in  an  amplitude  that  is  twice  the  projected  rotation.
Therefore, M4 exhibits  rotation at $0.9\pm0.1$\,km\,s$^{-1}$, with an
approximate  axis  of  rotation  of  PA=$70^\circ-250^\circ$,  M12  at
$0.15\pm0.1$\,km\,s$^{-1}$,  with an approximate  axis of  rotation of
PA=$40^\circ-220^\circ$ (although this  is effectively consistent with
no  rotation),   NGC  288  at   $0.25\pm0.15$\,km\,s$^{-1}$,  with  an
approximate axis  of rotation of PA=$0^\circ-180^\circ$,  and NGC 6752
shows  no rotation to  a level  of $0.2$\,km\,s$^{-1}$.   Our rotation
measurement  for M4 agrees  well with  that by  \cite{Peterson95}, who
quoted an  amplitude of  $0.9\pm0.4$\,km\,s$^{-1}$ with an  axis along
the line  PA=$100^\circ-280^\circ$, although only for  the inner $15'$
($\sim r_t/2$, about $1/3$ of the radius of our most distant member).

For all clusters, we corrected the individual stellar velocity data
for the measured rotation before calculating the velocity dispersions
and M/L$_{\rm V}$ profiles.

\subsection{Velocity Dispersions}\label{veldisp}

The systemic velocities of each cluster were measured using a Markov
Chain Monte Carlo (MCMC) method \citep{Gregory05}, taking into account
the individual velocity uncertainties on the stars, and providing the
systemic velocities with associated uncertainties. A simple
combination of the stellar velocities in each bin can provide a
measure of systemic velocity and dispersion, although this does not
take into account the individual velocity uncertainties. To fully
incorporate these, we used a Bayesian MCMC-based analysis to provide a
measure of realistic uncertainties of the velocity properties as a
function of radius. Our systemic velocities agree very well with
those from the literature ($V_r$ for all 10 clusters from this project
are shown in Table \ref{metaltable}), except for NGC 6144.  Our survey
did not sample this cluster well (19 stars were found to be members),
however, only 7 stars were analysed by \cite{Geisler95} which may
explain the discrepancy between the two values.  The literature values
are taken from \cite{Suntzeff93} (M4 and M12), \cite{Rutledge97} (NGC
288 and NGC 6752) and \cite{Geisler95} (NGC 6144).

The velocity dispersions of our samples were calculated for annular
bins centred on the cluster, each containing a similar number of stars
(M4 $\approx20$, M12 $\approx30$, NGC 288 $\approx20$ and NGC 6752
$\approx40$), centred on the cluster. The MCMC method described above
was used to determine the dispersion in each bin, with the resulting
velocity dispersion profiles overplotted with the best-fitting
line-of-sight \cite{Plummer11} model:

\begin{equation}\label{veldispeqn}
\sigma^2(R)=\frac{\sigma_0^2}{\sqrt{(1+R^2/r_s^2)}}.
\end{equation}

\noindent Here, $\sigma_0$ is the central velocity dispersion and
$r_s$ is the Plummer scale radius \cite[note that for Plummer models
the \new{unprojected} half-mass radius is $\approx1.305$ times the
scale radius\new{, however, for projected Plummer models such as in
this paper $r_s$ is equivalent to the projected half-mass
radius};][]{Haghi09}.  The Plummer model is advantageous for our
analysis because it is monotonically decreasing, so any flattening of
the profiles would be discernible.  It also allows for the calculation
of the total mass of the cluster from the central velocity dispersion
($\sigma_0$) and $r_s$ via (see \citealt{Dejonghe87} for a discussion
of Plummer models and their application):

\begin{equation}\label{totmass}
M_{tot} = \frac{64\sigma_0^2r_s}{3\pi G}.
\end{equation}

\noindent Note that this model assumes the velocity distributions of
the clusters are isotropic. How this assumption affects the overall
conclusions of this paper regarding MOND and dark matter within GCs is
far from obvious, and is a very complex problem which is beyond the
scope of the current study \cite[see][and references therein, for
detailed discussions of the problem of isotropy in
GCs]{Spurzem05,Giersz06,Kim08}. However, the anisotropies in each of
the ten clusters analysed in Papers I, II and the current study are
likely to vary greatly due to the large variation in Galactocentric
distances and rotational velocities. Because all our clusters exhibit
similar kinematic morphologies, with the possible exception of M4
(Section \ref{tidalheating}), the anisotropy issue seems to have
little impact on our results.

\begin{figure*}
  \begin{centering}
  \includegraphics[angle=-90,width=0.48\textwidth]{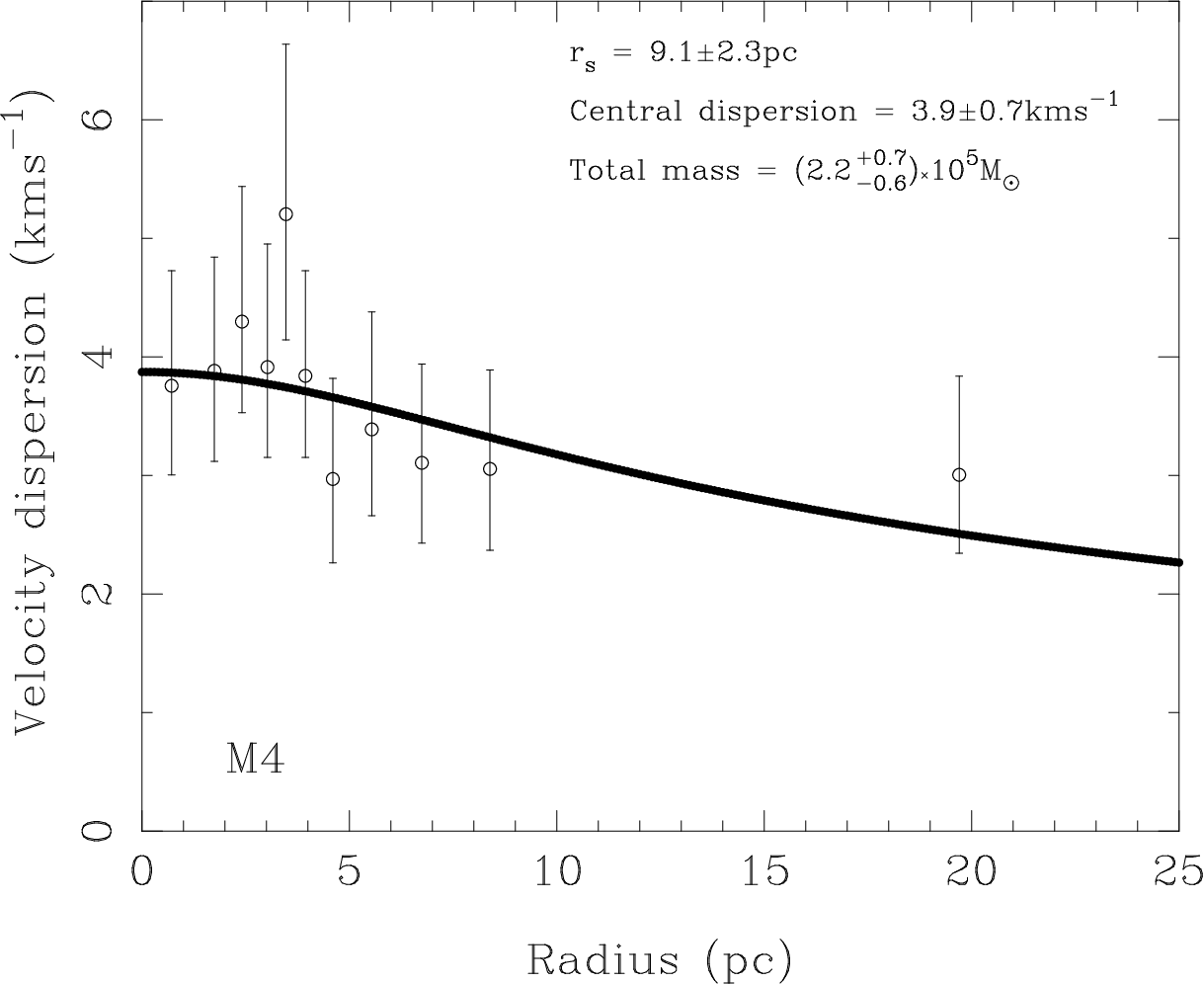}
  \includegraphics[angle=-90,width=0.48\textwidth]{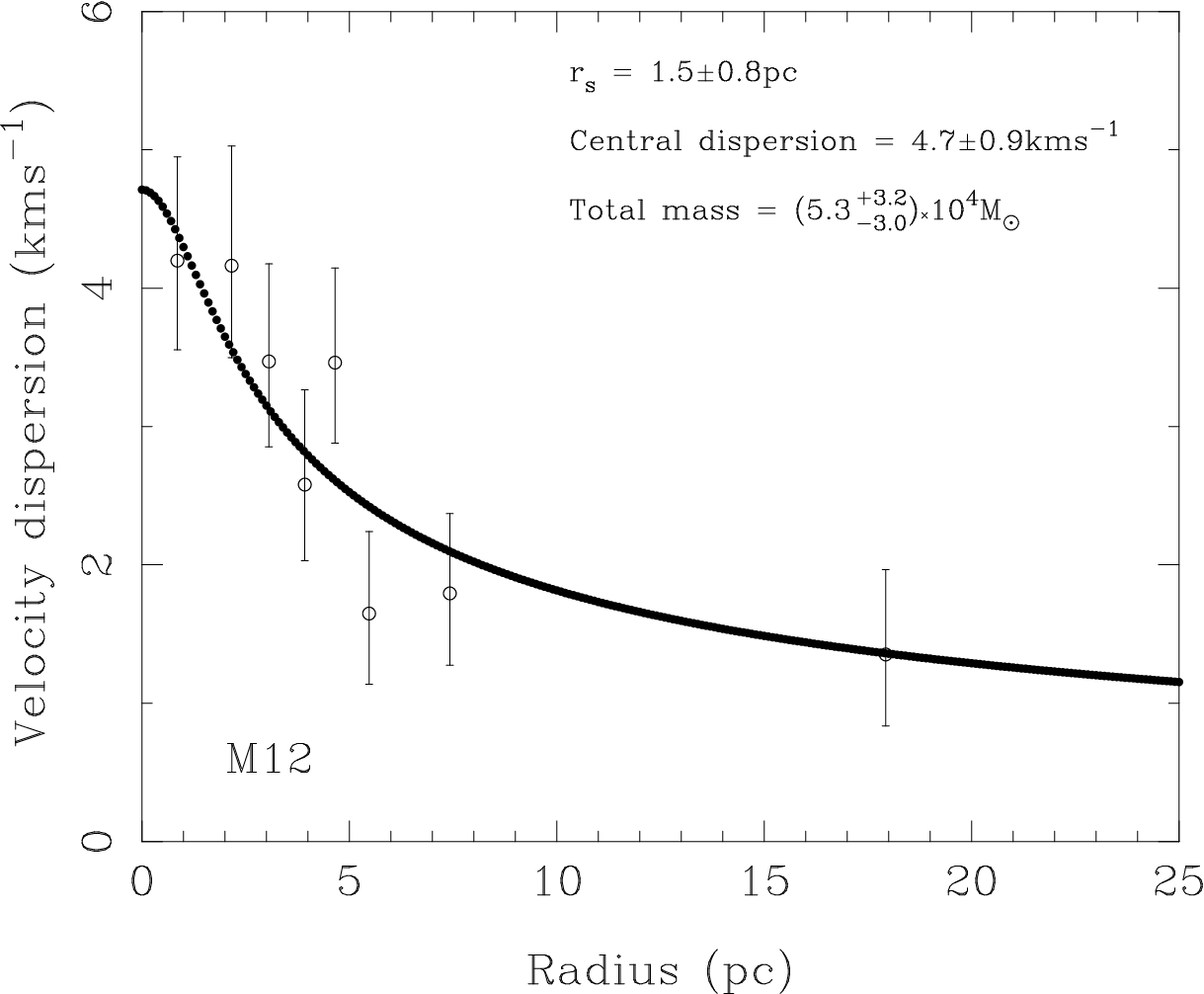}
  \includegraphics[angle=-90,width=0.48\textwidth]{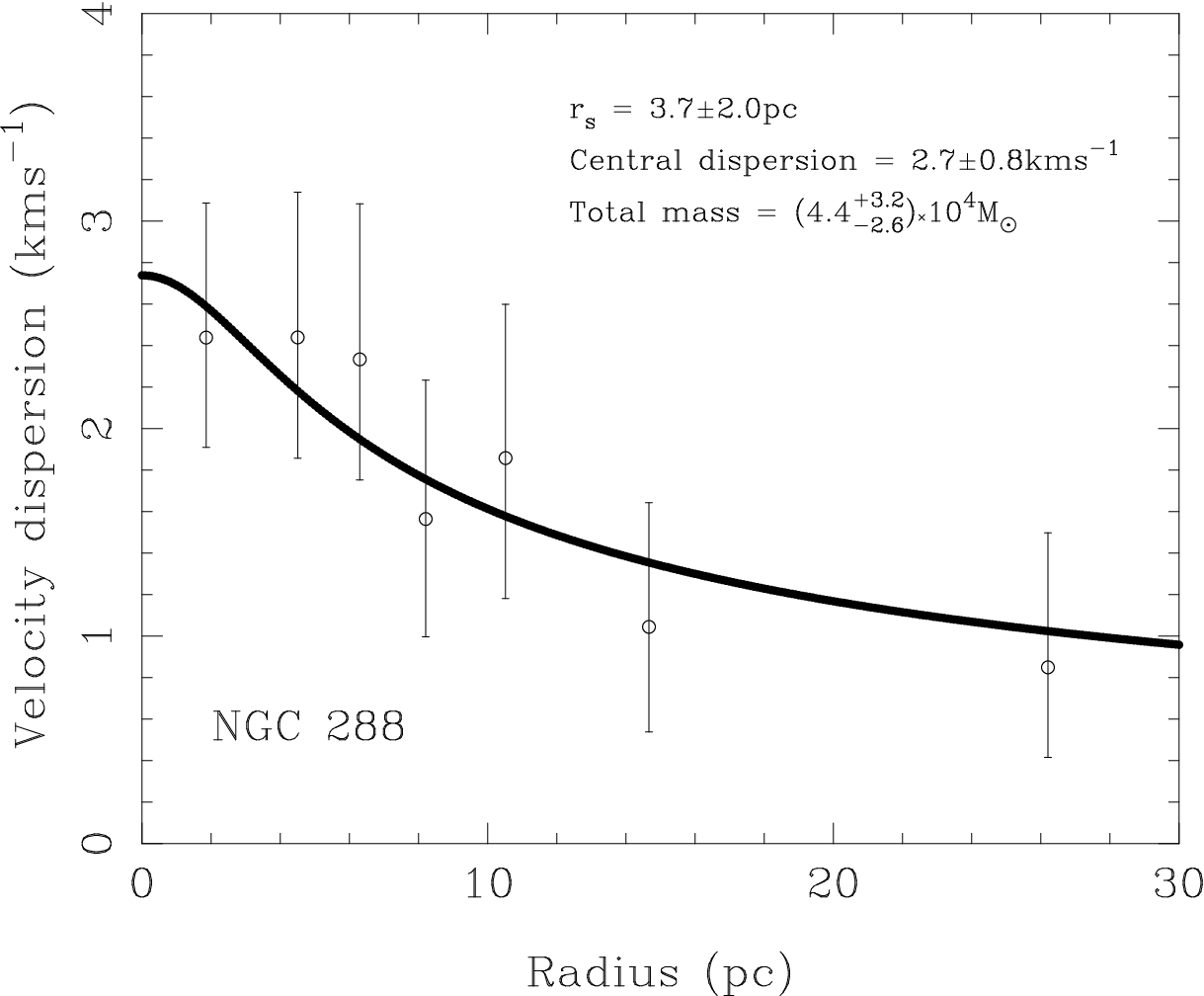}
  \includegraphics[angle=-90,width=0.48\textwidth]{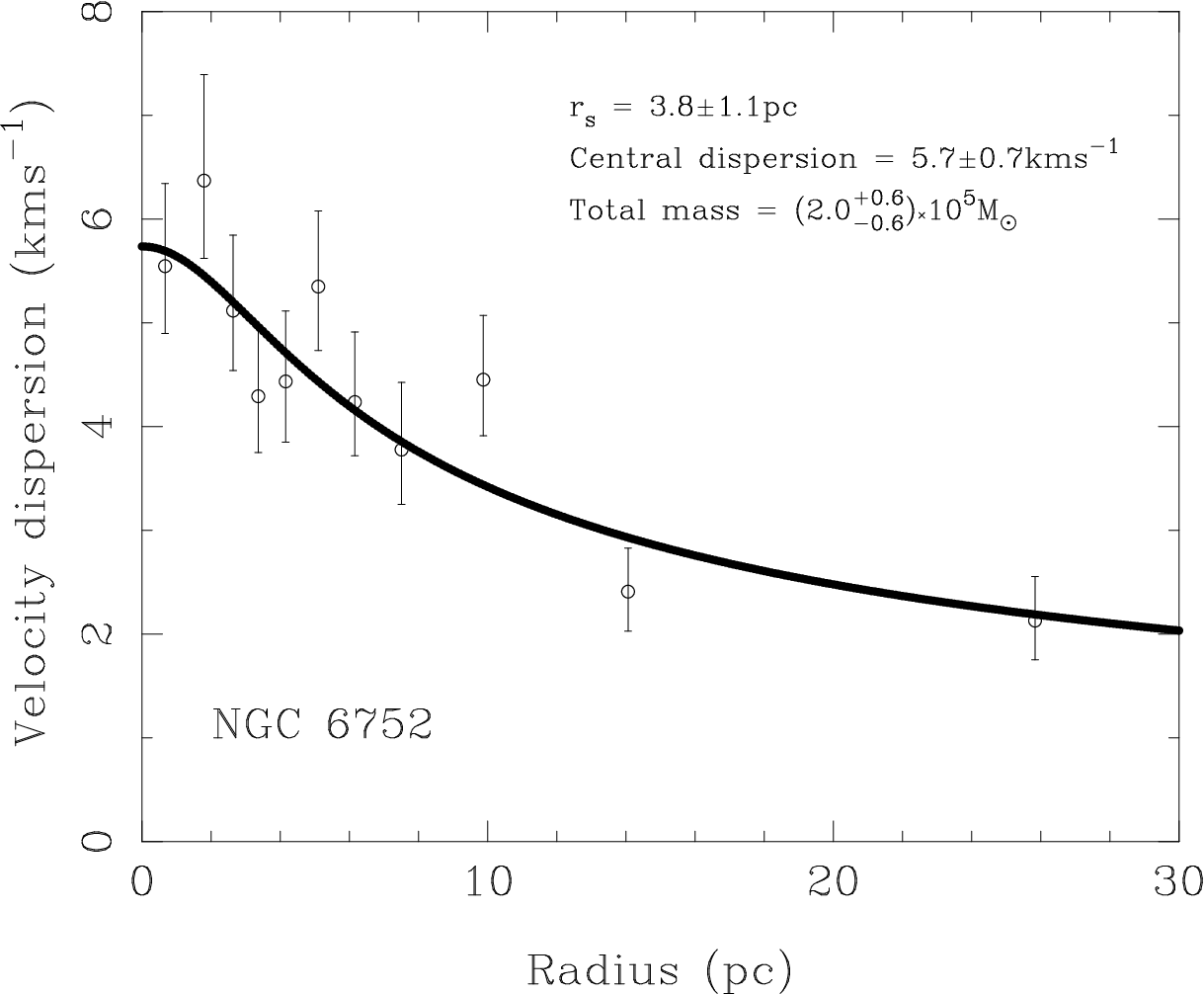}
  \caption{Velocity  dispersion  profiles  of  each cluster  from  the
  current   paper.  The  best   fitting  \citet{Plummer11}   model  is
  overplotted  and the  derived scale  radius, central  dispersion and
  total mass is shown in each panel.}
  \label{dispersions}
  \end{centering}
\end{figure*}

We have chosen a Plummer model that does not include a tidal cutoff,
over a more sophisticated Plummer model which includes a limiting
tidal radius, because the tidal radii of many GCs are not well known
(e.g. NGC 6752 and 47 Tucanae; see Section \ref{r_t6752} and Paper II,
respectively). Indeed, for many clusters we have found cluster members
well outside literature values of $r_t$. This means that our model
includes velocity dispersion information outside the tidal radii of
our clusters. We have, therefore, not removed any velocity information
for $R>r_t$, which would reduce the accuracy of the model at large
radii where $a<a_0$.

The velocity dispersion profiles, along with the total masses, scale
radii and central velocity dispersions are presented in Figure
\ref{dispersions}.  Except for M4, our mass estimates agree well with
other studies \cite[e.g.][]{Meylan89,Pryor93,Kruijssen09}, none of
whom used Plummer models to calculate their estimates. For M4, we find
a total mass about twice that of those studies [although this is
reduced to a $\sim69$\% difference if the extremes of the
uncertainties of \cite{Kruijssen09} and the current paper are taken].
Despite the Plummer profile fit being within the uncertainties, it is
apparent that the outer five bins of M4 have nearly the same measured
velocity dispersions.  This increases the value of $r_s$ in the
fitting of the profile, which is used to calculate the total mass
(Equation \ref{totmass}), which, in turn, leads to an inflated mass
estimate.  We attribute this apparent flattening of the dispersion
profile to tidal heating (see Section \ref{tidalheating}).

\begin{figure*}
  \begin{centering}
  \includegraphics[angle=-90,width=0.4\textwidth]{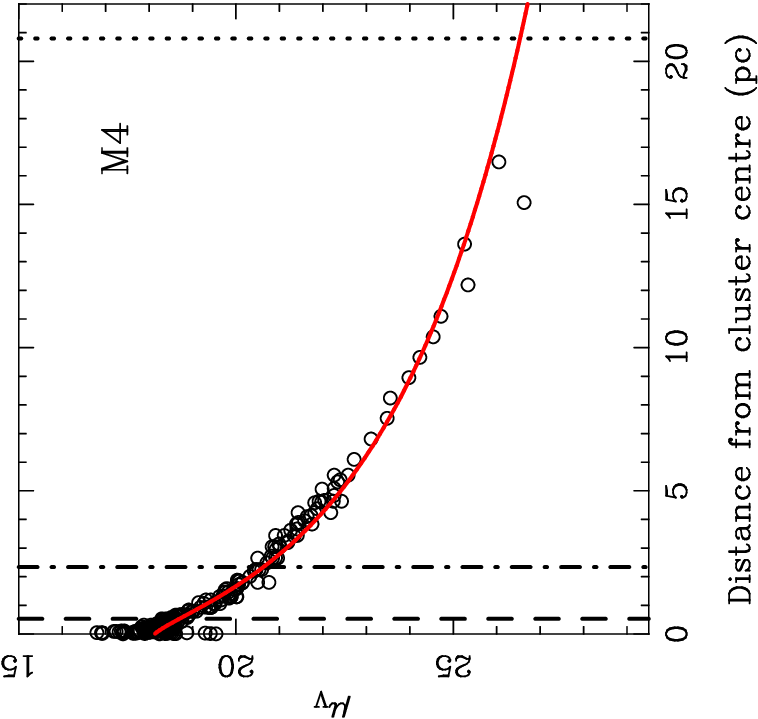}
  \includegraphics[angle=-90,width=0.4\textwidth]{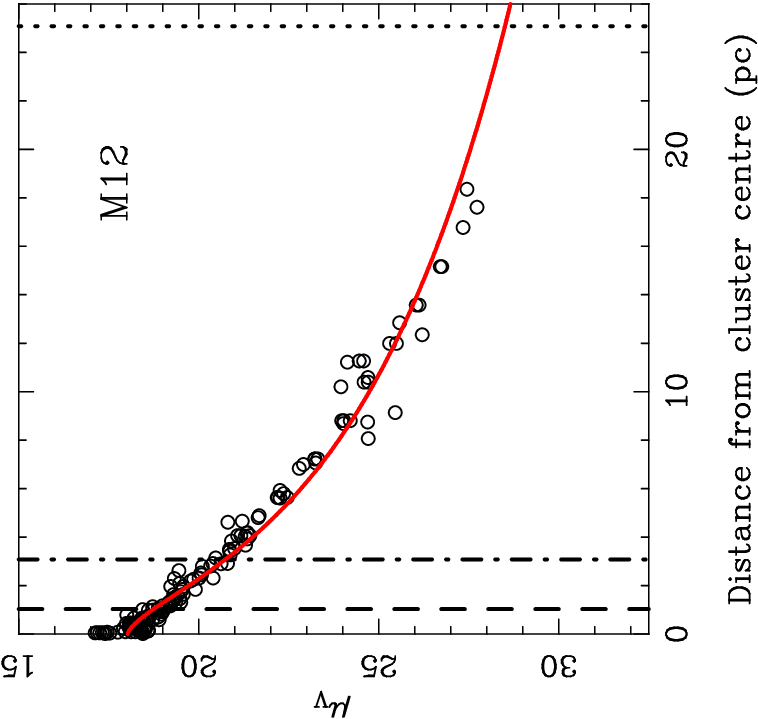}
  \includegraphics[angle=-90,width=0.4\textwidth]{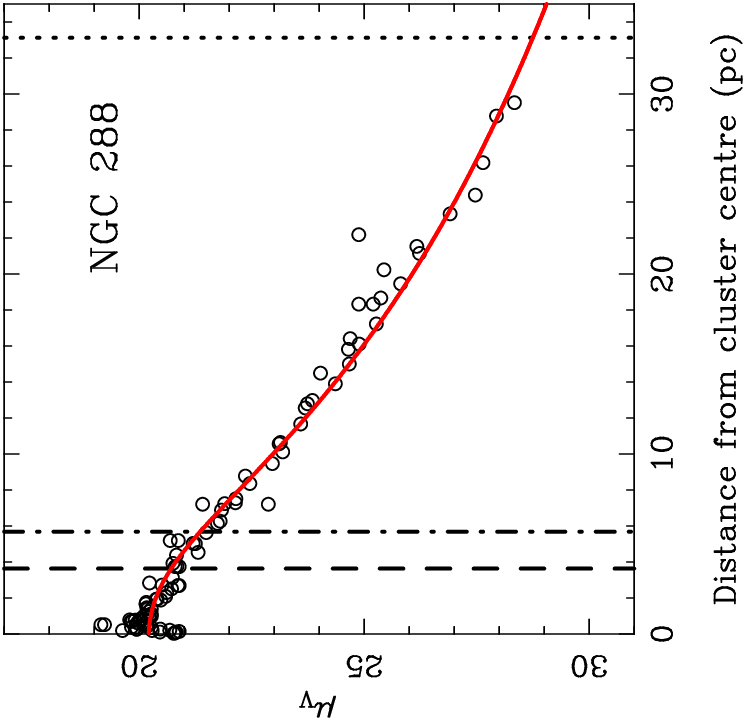}
  \includegraphics[angle=-90,width=0.4\textwidth]{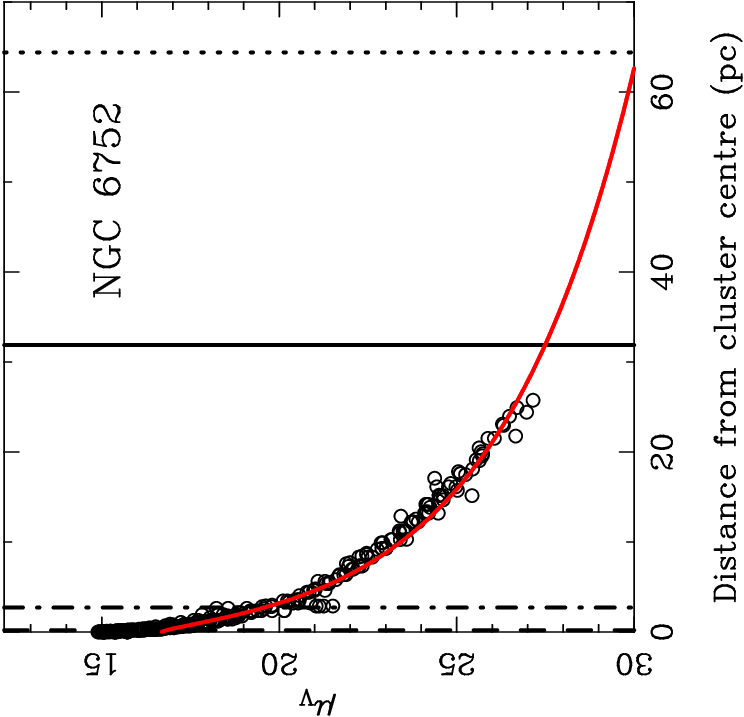}
  \caption{Surface brightness data by \citet{Trager95} overplotted
with best fitting Plummer profiles \new{[note that these are fitted
independently to the kinematic profiles in Section \ref{veldisp} and
lead to a different value of $r_s$ (denoted $r_{s,L}$), see
text]}. The vertical lines represent the core radius (dashed),
half-mass radius (dot-dashed) and tidal radius (dotted) from
\citet{Harris96}.  The solid vertical line in the lower right panel
(NGC 6752) is the tidal radius we derived in Section \ref{r_t6752}.}
  \label{tragerSB}
  \end{centering}
\end{figure*}

Except for M4, none of the clusters discussed here have shown the
apparent flattening of the velocity dispersion profiles reported by
\citet{Scarpa03,Scarpa07a,Scarpa07b}, indicating that neither a
significant DM component, nor a modified theory of gravity, is
required to explain their kinematic properties. This corroborates
earlier results for 47 Tuc, M22, M30, M53, M55 and M68 in Papers I and
II, and similar conclusions are drawn by \cite{Sollima09} for
$\omega$\,Centauri, by \cite{Jordi09} for Pal 14 and by
\cite{Sollima10} for MOND theories in general.  Through studies such
as these, it is becoming increasingly apparent that neither DM, nor
modified gravity theories, are necessary to explain the internal
kinematics of GCs.

\subsection{Projected Mass-to-Light Profiles}\label{M/L}

Dark matter causes larger stellar accelerations, and hence higher
maximal stellar velocities, therefore, a good indication of whether a
pressure-supported object like a GC is dark matter dominated is to
measure its mass-to-light ratio. To calculate the M/L$_{\rm V}$ for
our clusters we have used the surface brightness data by \cite[][see
Figure \ref{tragerSB}]{Trager95} to which projected Plummer surface
brightness profiles:

\begin{equation}\label{LR}
I(R)=\frac{L_{\rm tot}}{\pi}\frac{r_{s,L}^2}{(r_{s,L}^2 + R^2)^2}
\end{equation}

\noindent have been fitted (Figure \ref{tragerSB}). These Plummer
profiles were converted to solar luminosities per square
parsec. Projected mass density profiles \citep{Dejonghe87}:

\begin{equation}\label{MR}
\Sigma(R) = \frac{M_{\rm tot}}{\pi}\frac{r_s^2}{(r_s^2 + R^2)^2},
\end{equation}

\noindent in units of solar masses per square parsec, were then
divided by the surface brightness profiles to produce radial
mass-to-light profiles. The Plummer fits to the surface brightness
data do not include tidal cutoffs, for the reasons discussed in
Section \ref{veldisp}. Note that the kinematic and surface brightness
models have been fitted independently, \new{leading to two independent
values of $r_s$. The scale radius from the Plummer fits to the
luminosity profiles (Equation \ref{LR} and Figure \ref{tragerSB}) is
denoted $r_{s,L}$; throughout this paper $r_s$ represents the value
obtained from the cluster kinematics (Equation
\ref{veldispeqn}). These separate fits} allow a radial,
\new{projected}, M/L$_{\rm V}$ profile to be calculated:

\begin{equation}
{\rm M/L_V} = \frac{M_{\rm tot}}{L_{\rm tot}}\left(\frac{r_s}{r_{s,L}}\right)^2\left(\frac{r_{s,L}^2+R^2}{r_s^2+R^2}\right)^2.
\end{equation}

\begin{figure*}
  \begin{centering}
  \includegraphics[angle=-90,width=0.45\textwidth]{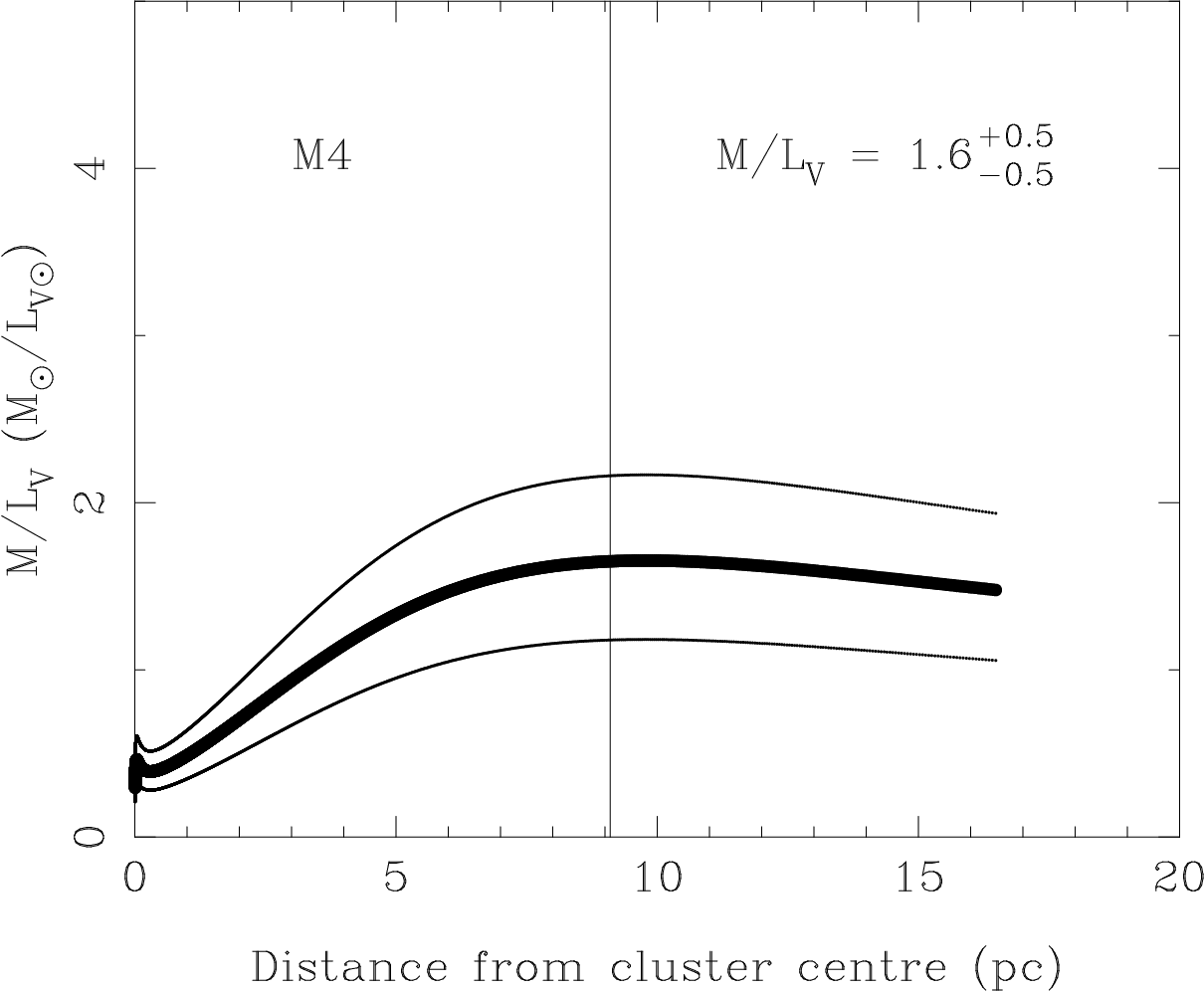}
  \includegraphics[angle=-90,width=0.45\textwidth]{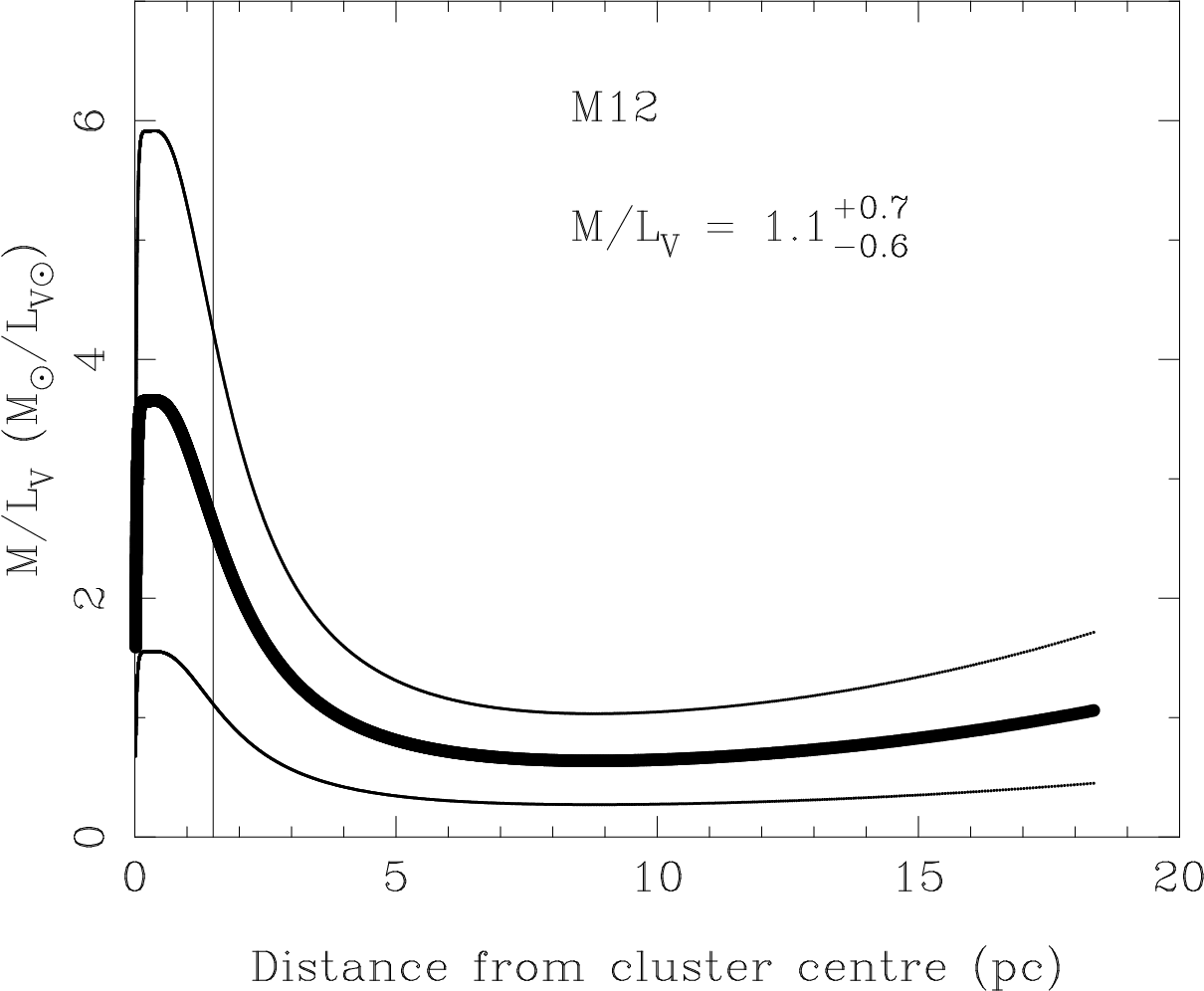}
  \includegraphics[angle=-90,width=0.45\textwidth]{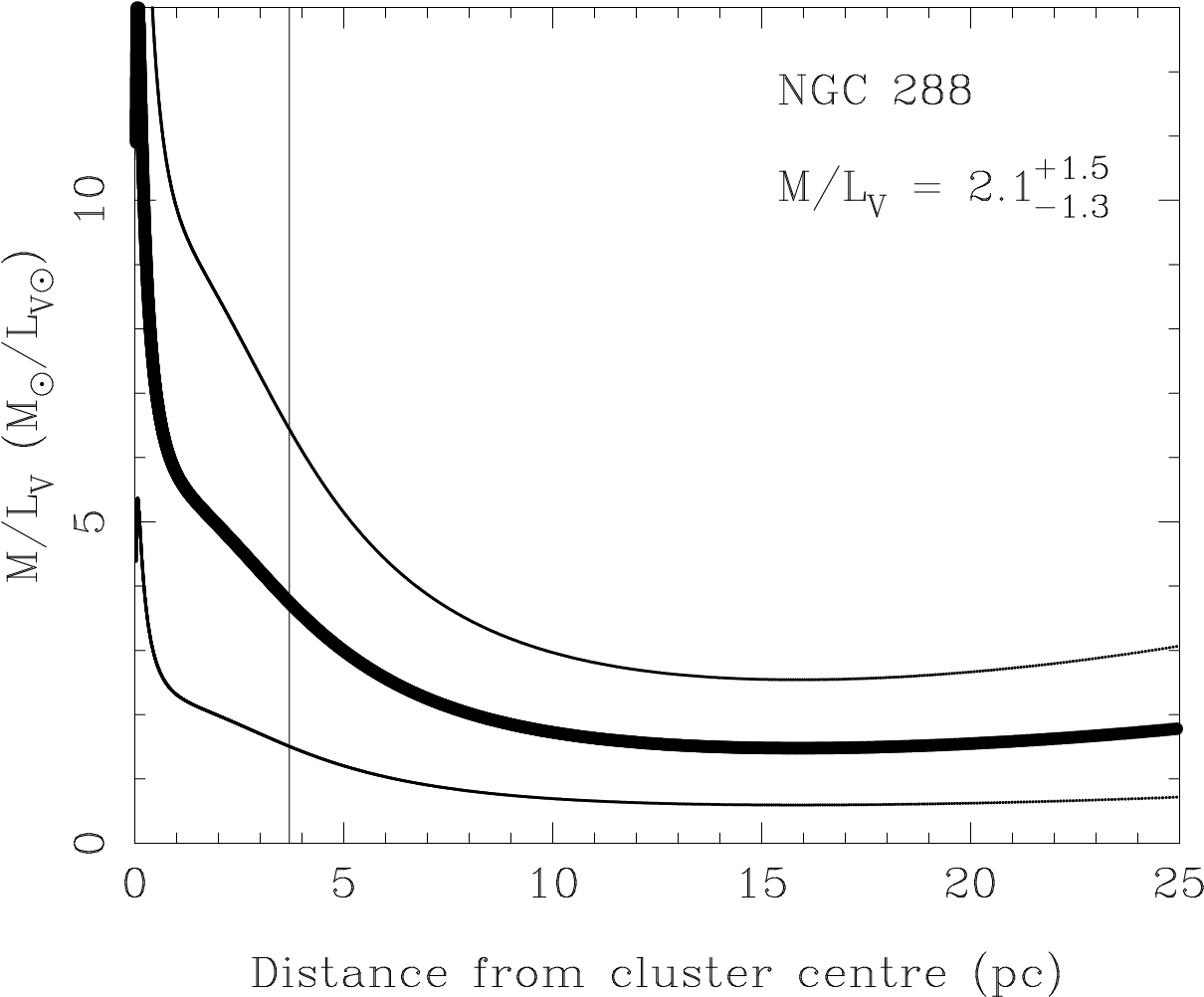}
  \includegraphics[angle=-90,width=0.45\textwidth]{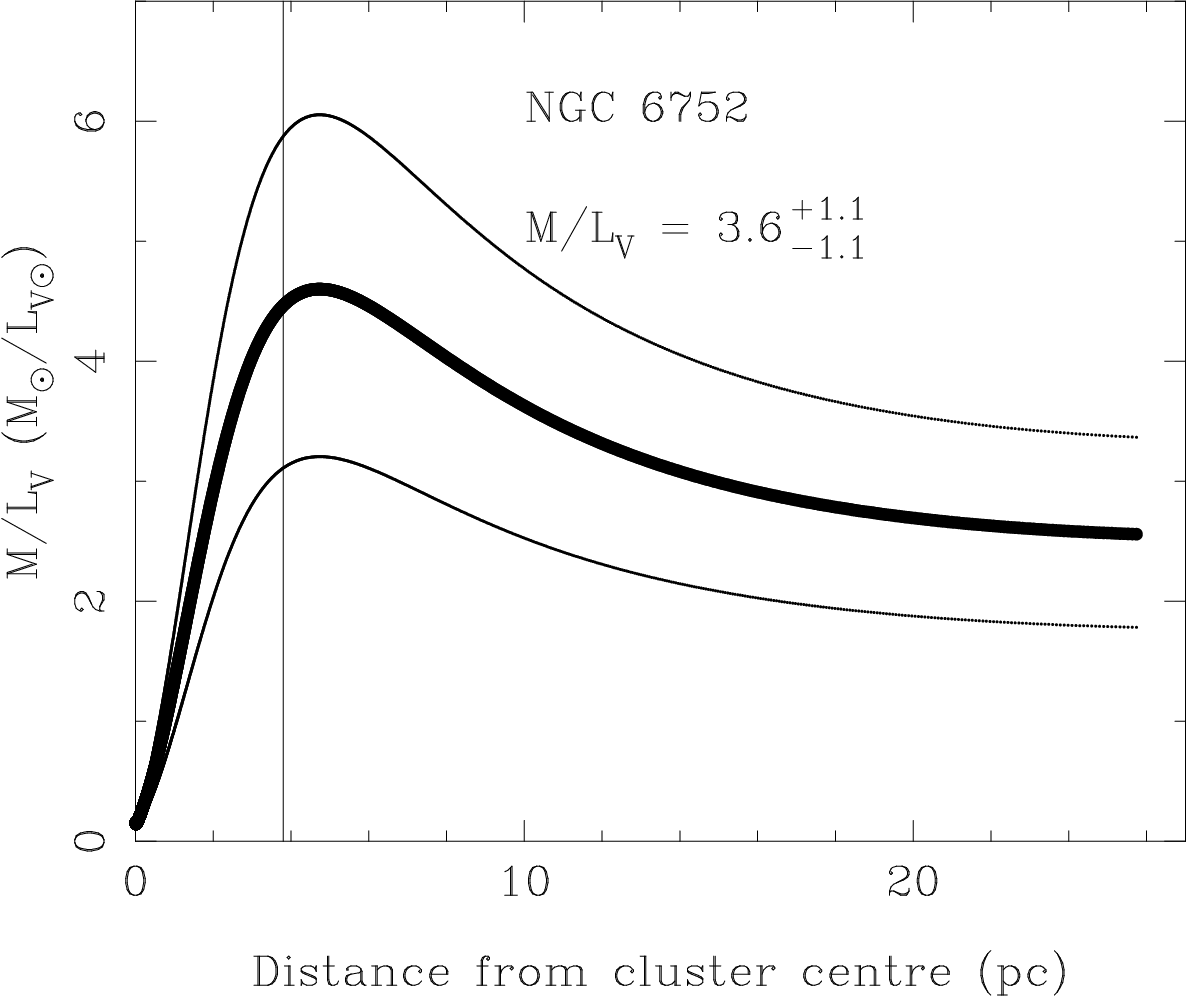}
  \caption{Mass-to-light profiles the the four clusters analysed in
    the current paper.  The thick line is the calculated M/L$_{\rm
    V}$, the thin lines are their uncertainties and the vertical line
    is the value of $r_s$. The quoted M/L$_{\rm V}$ value is only
    calculated for $R>r_s$.  None of these clusters have M/L$_{\rm
    V}\gg1$, furthering the argument that dark matter is not
    dominant. The deviation from a purely Plummer profile (Figure
    \ref{tragerSB}) has the effect of decreasing the M/L$_{\rm V}$ of
    NGC 6752 by $\sim0.8$ from its calculated value.}
  \label{MLfig}
  \end{centering}
\end{figure*}

The M/L$_{\rm V}$ profiles and mean values are shown in Figure
\ref{MLfig}; the thick line is the calculated M/L$_{\rm V}$ and the
thin lines are their uncertainties. The M/L$_{\rm V}$ profiles at
small radii deviate significantly from the mean. We interpret this as
being due to the uncertainty in the measured luminosities and
kinematics near the cores of the clusters. Crowding and confusion
effects are inherent in luminosity and kinematic measurements of dense
stellar fields, such as those near the cores of GCs; it is clear in
Figure \ref{tragerSB} that there is a large spread of surface
brightness measurements near the cores of all four clusters. Because
of this uncertainty in core luminosities and kinematics, our
mean M/L$_{\rm V}$ values were calculated for $R>r_s$ to ensure
that these effects were removed.

Dark-matter-dominated dynamical systems (e.g.  elliptical and dwarf
galaxies), exhibit high mass-to-light ratios (M/L$_{\rm V}\gtrsim10$),
whereas Ultra Compact Dwarfs, which have the same velocity
dispersion--luminosity relation as GCs
\cite[][]{Hasegan05,Evstigneeva07}, show no evidence for DM for
M/L$_{\rm V}\lesssim5$.  None of our clusters have M/L$_{\rm V}\gg1$,
therefore DM cannot dominate, and because none have M/L$_{\rm V}>5$ we
see no need for any DM component.  Similar conclusions were reached
for all six clusters studied in Papers I and II. All results from this
project indicate strongly that, in general, GCs do not contain large
quantities of DM. We thought it important to mention that
\cite{Baumgardt08} have shown that dynamically more evolved GCs
exhibit lower M/L$_{\rm V}$ values, so the larger M/L$_{\rm V}$ of NGC
6752 should indicate that this cluster is dynamically
``young''. However, this is in direct contradiction with the current
understanding of NGC 6752 having a collapsed core
\cite[][]{Rubenstein97}.

It is interesting to note that \cite{Ferraro03} discussed the
possibility of a large M/L$_{\rm V}$ value ($\sim6-7$) for the inner
0.08\,pc of NGC 6752, because of the observed accelerations of
millisecond pulsars near the core. Since we do not claim any knowledge
of the M/L$_{\rm V}$ at those radii, this indeed remains a
possibility, should the M/L$_{\rm V}$ increase further from $r_s$
toward the core.  This intriguing possibility should be pursued by
extracting more information on the kinematics of the core of NGC 6752,
particularly in light of a newly recognised correlation that may be
useful for accurately estimating the masses of GC cores
\cite[][]{Leigh09}.

\begin{table*}
\begin{center}
\caption{All parameters derived from the kinematics of the
clusters in this project in order of decreasing metallicity (see
text, \citetalias{Lane09} and \citetalias{Lane10a} for literature
comparisons), as well as the tidal radius from \citet{Harris96}
and the estimated acceleration, due to the cluster, for the most
distant cluster member. From left to right the columns are: cluster
name, [Fe/H], systemic radial velocity, M/L$_{\rm V}$, total cluster
mass, rotational velocity, Plummer scale radius, central velocity
dispersion, tidal radius and acceleration due to the cluster.
Some parameters were not calculated due to low sampling, and 47 Tuc
and M55 do not have calculated values for [Fe/H] because they were
used as calibrators (see Paper II).  $V_r$, $V_{\rm rot}$ and
$\sigma_0$ are in km\,s$^{-1}$, $M_{\rm tot}$ is in
${\rm10^5\,M_\odot}$, $r_s$ and $r_t$ are in pc and $a$ is
in m\,s$^{-2}$. The $r_t$ value for NGC 6752 in parentheses is
that derived in Section \ref{r_t6752}.}\label{metaltable}
\begin{tabular}{@{}cccccccccc@{}}
\hline
\hline
Cluster & [Fe/H] & $V_r$ & M/L$_{\rm V}$ & ${M}_{\rm tot}$ & $V_{\rm
  rot}$ & $r_s$ & $\sigma_0$ & $r_t$ & $a$\\
\hline
47 Tuc & --\hspace{-0.24cm} & $-16.9\pm0.2$ & $4.6^{+0.6}_{-0.6}$ &
$11^{+1}_{-1}$ & $2.2\pm0.2$ & $7.8\pm0.9$ & $9.6\pm0.6$ & 56.1 &
$2.3\times10^{-11}$\\
Kron 3 & $-1.05\pm0.10$ & -- & -- & -- & -- & -- & -- & -- & --\\
M4 & $-1.16\pm0.13$ & $71.5\pm0.3$ & $1.6^{+0.5}_{-0.5}$ &
$2.2^{+0.7}_{-0.6}$ & $0.9\pm0.1$ & $9.1\pm2.3$ & $3.9\pm0.7$ & 20.8 &
$3.8\times10^{-11}$\\
NGC 288 & $-1.24\pm0.13$ & $-45.1\pm0.2$ & $2.1^{+1.5}_{-1.3}$ &
$0.44^{+0.32}_{-0.26}$ & $0.25\pm0.15$ & $3.7\pm2.0$ & $2.7\pm0.8$ &
33.1 & $6.2\times10^{-12}$\\
Sgr & $-1.40\pm0.50$ & -- & -- & -- & -- & -- & -- & -- & --\\
NGC 121 & $-1.50\pm0.10$ & -- & -- & -- & -- & -- & -- & -- & --\\
M12 & $-1.50\pm0.13$ & $-41.0\pm0.2$ & $1.1^{+0.7}_{-0.6}$ &
$0.53^{+0.32}_{-0.30}$ & $0.15\pm0.1$ & $1.5\pm0.8$ & $4.7\pm0.9$ &
25.1 & $1.1\times10^{-11}$\\
NGC 6752 & $-1.62\pm0.15$ & $-26.2\pm0.2$ & $3.6^{+1.1}_{-1.1}$ &
$2.0^{+0.6}_{-0.6}$ & nil & $3.8\pm1.1$ & $5.7\pm0.7$ & 64.4
($31.9\pm2.0$) & $2.7\times10^{-11}$\\
M22 & $-1.78\pm0.15$ & $-144.9\pm0.3$ & $4.7^{+1.7}_{-1.7}$ &
$3.3^{+1.2}_{-1.1}$ & $1.5\pm0.4$ & $4.5\pm1.5$ & $6.8\pm0.9$ & 27.0 &
$6.4\times10^{-11}$\\
M55 & --\hspace{-0.24cm} & $174.8\pm0.4$ & $2.0^{+0.9}_{-0.8}$ &
$1.4^{+0.5}_{-0.5}$ & $0.25\pm0.09$ & $11.7\pm4.2$ & $2.7\pm0.5$ &
25.1 & $1.9\times10^{-11}$\\
M53 & $-1.99\pm0.10$ & $-62.8\pm0.3$ & $6.7^{+1.9}_{-1.7}$ &
$5.2^{+1.5}_{-1.4}$ & nil & $17.2\pm3.8$ & $4.4\pm0.9$ & 112.6 &
$5.8\times10^{-12}$\\
M68 & $-2.06\pm0.15$ & $-94.9\pm0.3$ & $1.9^{+1.0}_{-0.8}$ &
$0.57^{+0.29}_{-0.24}$ & $0.6\pm0.4$ & $6.4\pm2.0$ & $2.4\pm0.9$ &
90.0 & $1.9\times10^{-12}$\\
M30 & $-2.16\pm0.15$ & $-184.4\pm0.2$ & $1.5^{+0.9}_{-0.8}$ &
$0.90^{+0.51}_{-0.48}$ & nil & $2.3\pm1.2$ & $5.0\pm0.9$ & 42.7 &
$8.1\times10^{-12}$\\
NGC 6144 & --\hspace{-0.24cm} & $196.6\pm0.8$ & -- & -- & -- & -- & --
& 82.2 & --\\
\hline
\end{tabular}
\end{center}
\end{table*}

\subsubsection{Correlation Between M/L$_{\rm V}$ and Luminosity}\label{47T}

\begin{figure}
  \begin{centering}
  \includegraphics[angle=-90,width=0.45\textwidth]{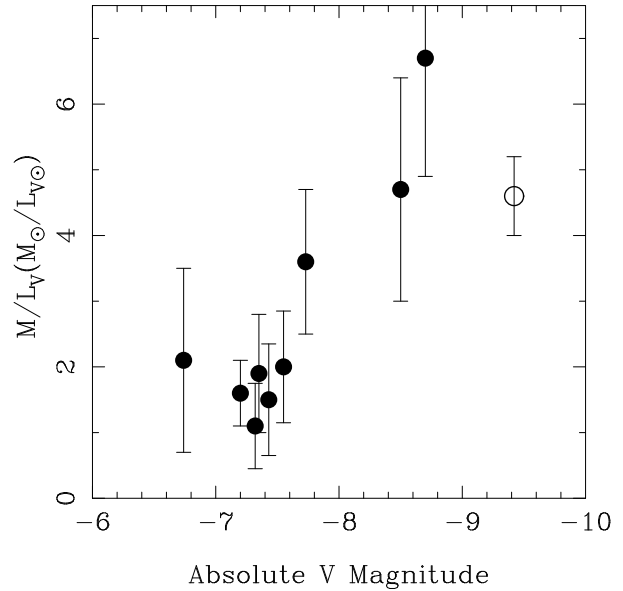}
  \caption{M/L$_{\rm V}$ from this study vs absolute V band
  magnitude \citep[taken from][]{Harris96} for all 10 clusters.  Note
  the trend to higher M/L$_{\rm V}$ with increased
  luminosity. The open circle is 47 Tuc.}
  \label{M/Lvsmass}
  \end{centering}
\end{figure}

47 Tuc is well known as having a bimodal distribution of various
line strengths in both MS and giant stars
\cite[e.g.][]{Norris79,Cannon98,Harbeck03}.  The most popular
explanation for this phenomenon is that the cluster has undergone
multiple episodes of star formation.  In Paper II we reported a rise
in the velocity dispersion of 47 Tuc for $R\gtrsim r_t/2$, which we
interpreted as evidence for evaporation.  Here we present an
alternative scenario for this signature: a past merger, which explains
both the rise in velocity dispersion and the bimodality in line
strengths, as well as its anomalously high luminosity compared
with its M/L$_{\rm V}$ (see Figure \ref{M/Lvsmass}).

The clusters from this project exhibit a clear trend between M/L$_{\rm
V}$ and total luminosity \cite[Figure \ref{M/Lvsmass}; a trend which
is generally attributed to low-mass star depletion, e.g.][and
references therein]{Kruijssen09.2}, with 47 Tuc being an apparent
outlier.  Since there is no reason to think that 47 Tuc has a large DM
component, a simple explanation is that 47 Tuc merged with another
object of similar metallicity in its past.  This merger would have
increased the total luminosity of the cluster without altering its
M/L$_{\rm V}$ significantly and caused bimodalities in both line
strengths and velocities.  If there has not yet been enough time for
the two populations to mix thoroughly, this may be the cause of the
observed rise in velocity dispersion.  A detailed analysis of this
scenario was presented by \cite{Lane10b}.

\section{Evidence for Tidal Heating}\label{tidalheating}

When attempting to determine the reality of MOND using GCs, it is
generally agreed that GCs at large distances from the Galaxy are most
useful because the external acceleration imparted by the Galactic
tidal field is below $a_0$ \cite[][and references
therein]{Sollima10}. An important question to ask, then, is at
what Galactocentric distance does the external field become
``negligible''?  Furthermore, does the shape of the Halo have an
effect on the GC dynamics, whereby distant GCs in certain regions of
the Halo are affected above this threshold?  One clear theoretical
prediction is that the external field should heat the external parts
of the GCs, thus increasing the velocity dispersion, up to, and
including, tidal destruction \cite[e.g. Pal
5;][]{Odenkirchen01}. This is especially true during Disc
crossings and at perigalacticon where tidal shocks strongly affect the
dynamics of the cluster for short periods. We present here an
examination of our data as an analysis of the Galactic tidal
field.

\begin{figure}
  \begin{centering}
  \includegraphics[angle=-90,width=0.45\textwidth]{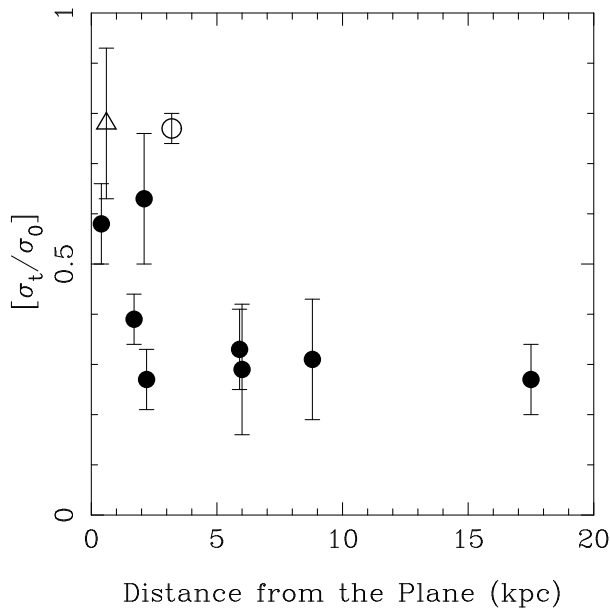}
  \caption{The ratio between velocity dispersions at the tidal radius
  and the core vs distance from the Plane \citep[taken
  from][]{Harris96} for the 10 clusters.  Note the trend toward higher
  values of [$\sigma_t/\sigma_0$] toward the Galactic plane.
  The open circle is 47 Tuc and the open triangle is M4.}
  \label{Zvsveldisp}
  \end{centering}
\end{figure}

Figure \ref{Zvsveldisp} shows how the ratio between the velocity
dispersions at the tidal radius and core varies with distance from the
Plane. The closer to the Plane, the greater the tidal effects from the
Galaxy, and the larger the ratio.  It is clear that for
$R_Z\gtrsim3$\,kpc the tidal effects of the Galaxy are essentially
equivalent at all radii. From this we can infer that the DM Halo
exerts the dominant tidal force for $R_Z\gtrsim3$\,kpc. Furthermore,
because the four clusters beyond $R=5$\,kpc (in increasing distance
these are M30, M68, NGC 288 and M53) are in different locations in the
Halo \cite[][]{Harris96} and on orbits with vastly different
orientations to the Halo \cite[][]{Allen06}, this indicates the
possibility of a non-triaxial DM Halo \cite[see][for a discussion of
the effects of halo triaxiality on the dynamics of GCs]{Penarrubia09}.
Note that this is not strong evidence for the shape of the dark Halo,
however, it is worth mentioning.

M4 has a fairly flat velocity dispersion profile in the outer regions
(Figure \ref{dispersions}). This cluster is also very close to the
Plane \cite[600\,pc;][]{Harris96} and has the largest value of
[$\sigma_t/\sigma_0$] of any of our clusters (Figure
\ref{Zvsveldisp}).  Furthermore, its 3D space velocity, with respect
to the Local Standard of Rest \cite[][]{Dinescu99}, indicates that it
is continuously interacting with the Disc because its orbit is nearly
Planar. This, combined with its low M/L$_{\rm V}$, strongly indicates
that tidal heating is the cause of the flattening of the velocity
dispersion in the outskirts of M4 rather than a substantial DM
component.

Figure \ref{Zvsveldisp} seems to indicate that GCs are tidally shocked
within $\sim3$\,kpc of the Plane, then cool down, and stay cool,
beyond that distance.  However, the orbital periods calculated by
\cite{Dinescu99} are about an order of magnitude shorter than the
relaxation times in the \cite{Harris96} catalogue for all of our
clusters.  Since the outskirts of GCs are much less dense than the
cores, two-body interactions could not cause the outer regions to cool
in less than the relaxation time of the cluster. In fact, no mechanism
known to the authors can account for this rapid cooling in the
outskirts of our GCs.  Based on the M/L$_{\rm V}$ and velocity
dispersion profiles of all 10 clusters, a large DM component is very
unlikely. It is also very unlikely that we are seeing a MONDian effect
in these outer regions because of the Newtonian nature of the velocity
dispersion profiles. Addressing this interesting problem is beyond the
scope of this paper, and more work is required to solve this
intriguing puzzle.

\section{Conclusions}

In the current paper we have analysed four GCs (M4, M12, NGC 288 and
NGC 6752) to determine their velocity dispersion and M/L$_{\rm V}$
profiles, bringing the total to 10 for this project. We have
included GCs that have external accelerations extending from above
$a_0$ down to $a_0$ and we find no deviation from our Plummer models
at lower accelerations. Therefore, we see no indication that DM, or a
modified version of gravitational theory, is required to reconcile
GC dynamics with Newtonian gravity. This adds to the growing
body of evidence that GCs are DM-poor, and that our understanding of
weak-field gravitation is not incomplete. Within the stated
uncertainties, the dynamics of all these clusters are well described
by purely analytic \citet{Plummer11} models, which indicates that
Newtonian gravity adequately describes their velocity dispersions, and
we observe no breakdown of Newtonian gravity at
$a_0\approx1.2\times10^{-10}$\,m\,s$^{-2}$, as has been claimed in
previous studies.

Despite this, we see the intriguing possibility of an unknown cooling
process in the outskirts of GCs; the external regions of our GCs seem
to cool much faster following tidal Disc shocks than the relaxation
time of the clusters.  Because it is highly unlikely that a MONDian
process, or a significant DM component, is the cause of this cooling
(based on our velocity dispersion and M/L$_{\rm V}$ profiles), further
work is required to solve this puzzle. Furthermore, the lack of tidal
heating events in the distant clusters provides some indication that
the dark Halo is not triaxial.

The Plummer model was also used to determine the total mass, scale
radius, and M/L$_{\rm V}$ profile for each cluster.  We find that none
of our clusters have M/L$_{\rm V}\gg1$, further evidence that DM does
not dominate.  We have produced M/L$_{\rm V}$ profiles, rather than
quoting a single value based on the central velocity dispersion and
central surface brightness.  This method is used because it
describes the M/L$_{\rm V}$ of the entire cluster, rather than only
its core. This is particularly important for post-core-collapsed GCs,
where crowding and confusion effects introduce significant
uncertainty into luminosity and kinematic measurements at small
radii.  Within the uncertainties, our estimated cluster masses all
match those in the literature except for M4, which we calculate to
have a total mass about twice that of the literature values.  The
reason for this discrepancy is that the tidally heated cluster has an
increased velocity dispersion in its outer regions, flattening the
Plummer fit, increasing the value of $r_s$, and therefore, increasing
the mass estimate.

Another important result from this study is the measured rotations of
our clusters.  Of the four clusters studied here, M4 and NGC 288 show
clear rotation, M12 may have some rotation, and NGC 6752 displays no
rotation signature.
% When 47 Tuc and NGC 6752 are placed on the low-mass
%extension of the M-$\sigma$ relation the black hole masses inferred do
%fit with other estimates of black holes in these clusters, and M22 may
%also contain a central compact mass of
%$\sim170^{+110}_{-75}$\,M$_\odot$. These M-$\sigma$ results, however,
%should be treated with caution.

Throughout this project we have found similar results for the dark
matter content, and Newtonian kinematics, of our 10 GCs, all at
varying distances from the Galactic centre and Disc, including
three that experience external accelerations due to the Galaxy of
$\sim a_0$. All data were acquired using the same instrument (AAOmega
on the Anglo-Australian Telescope), reduced using the same pipeline
({\tt 2dfdr}), and analysed in the same way.  This homogeneous
approach is vital to a large project such as this, to ensure all
systematics are accounted for in a similar fashion.  Because of all
these factors, our results from the three papers are $strongly$
indicative that the current picture of globular clusters being
dark-matter poor, and with dynamics explained by standard Newtonian
theory, is correct.

\section{Acknowledgements}

This project has been supported by the University of Sydney, the
Anglo-Australian Observatory, the Australian Research Council, the
Hungarian OTKA grant K76816 and the Lend\"ulet Young Researchers
Program of the Hungarian Academy of Sciences. RRL thanks Martine
L. Wilson for her support during the writing of this paper. The
authors thank Dean McLaughlin for his helpful comments.

\bsp

\label{lastpage}

\end{document}